# DFT based comparative study of the physical properties of *M*AlB (*M* = V, Ta, Mo, Nb) MAB compounds

Jahid Hassan, M. A. Masum, Ruman Ali, Md. Enamul Haque, S. H. Naqib*
Department of Physics, University of Rajshahi, Rajshahi 6205, Bangladesh
*Corresponding author; Email: salehnaqib@yahoo.com

## Abstract

MAB phases have appealing physical features that make them appropriate for a wide range of applications. Motivated by this, we present density functional theory (DFT) calculations of the structural, elastic, bonding, electronic band dispersion, acoustic behavior, phonon spectrum, various thermomechanical and optoelectronic properties of VAlB and TaAlB ternary borides for the first time. The computed ground state lattice parameters of both compounds are very consistent with experimental data. The formation enthalpy, elastic constants, and phonon dispersion calculations indicate that both compounds are chemically, mechanically, and dynamically stable, respectively. The physical parameters of VAlB and TaAlB are studied and compared with those of MoAlB and NbAlB MAB compounds. All of these MAB phases are elastically anisotropic, machinable, and brittle materials. Layered characteristics are reflected in structural and elastic properties in these compounds. The predicted electronic band structure, density of states, and optical properties of *M*AlB (*M* = V, Ta, Mo, Nb) structures show metallic characteristics. The conduction band near the Fermi level of VAlB and TaAlB compounds is dominated by V 3*d* and Ta 5*d* electronic states. The low-dispersive V 3*d* bands and Ta 5*d* bands contribute to the Fermi surface of VAlB and TaAlB compounds, respectively. Furthermore, several thermophysical characteristics for these compounds are evaluated, such as the Debye temperature, average sound velocity, melting temperature, heat capacity, lattice thermal conductivity, and so on. The complete mechanical and thermophysical analysis reveals that these MAB phase compounds have great structural stability, moderate hardness, high Debye temperature, and high melting temperature. The bonding nature has been investigated using charge density mapping and bond population analysis. The population analysis and charge density distribution map demonstrate the presence of both ionic and covalent bonds in these compounds, with the covalent bond playing the dominating role. We comprehensively investigated the reflectivity, absorption coefficient, refractive index, dielectric function, optical conductivity, and loss function of these ternary borides. These MAB compounds exhibit a degree of anisotropy in all optical the parameters. The optical absorption, reflectivity, and refractive index spectra of *M*AlB (*M* = V, Ta, Mo, Nb) demonstrate that they are good ultraviolet (UV) radiation absorbers with outstanding reflection properties, making them promising for application in optoelectronic devices.

## 1. Introduction

The current era is a "golden age of materials," as innovative approaches to materials science are being actively pursued to meet the growing demands of society to serve as structural, energy harvesting and energy storage systems, to foster better lifestyles [1]. The capacity of multifunctional materials to satisfy the ever-more-complex requirements of contemporary industry and technology is what gives them significance. They foster innovation in a variety of industries by enabling more effective, dependable, and sustainable solutions through the provision of several functionalities in a single material. The importance of multifunctional materials will grow as technology develops, creating new opportunities and uses. Our examined MAB compounds in this work show great promise as multifunctional materials with a broad variety of uses because of their special blend of mechanical, electrical, thermal, and structural characteristics. Because of their remarkable physical and chemical characteristics, compounds with nanolaminated crystal structures have been the subject of much research. These substances have high levels of hardness, thermal stability, electrical conductivity, and resistance to chemical changes because they are made of transition metal elements bound to boron, carbon, or nitrogen. The well-known and widely studied ternary carbides and nitrides, with the chemical composition $M_{n+1}AX_n$ (MAX), where n $\epsilon\{1,2,3\}$, M is an early transition metal, A is a A-group element (mostly IIIA and IVA) and X is C or N, are all layered hexagonal structures with space group *P6₃/mmc* (no. 194). These materials combine some properties of metals, such as good electrical and thermal conductivity, machinability, low hardness, thermal shock resistance and damage tolerance, with those of ceramics, such as high elastic moduli, high temperature strength, and oxidation and corrosion resistance [2,3]. As research continues, the range of applications for MAX compounds is expected to expand, further underscoring their importance in advancing technology and industrial capabilities.

Very recently, Yury Gogotsi and his coworkers reported that the etching of $Ti_3AlC_2$ that results in the selectively removing of the A element, usually Al, and produced $Ti_3C_2T_x$ [4] for the first time; a 2D nanomaterial that was both strong and flexible. They named the new material MXene (pronounced MAXene), that has demonstrated exceptional promise for a variety of uses, such as energy storage [5], nano-electronics [6], photochemical conversation [7] and electrocatalysis [8] etc. The 'MX' for the elements left after etching and the 'ene' as a reference to graphene, one of the most popular 2D materials. The aforementioned researches also inspired the related studies to search new MAX-like and MXene-like compounds [9–12]. Also, a class of ternary transition-metal borides (nanolaminated compounds) called MAB, with the general formula $(MB)_{2z}A_x(MB_2)_y$ (z = 1 – 2; x = 1 – 2; y = 0 – 2) [13], whose structures are composed of a transition M-B sublattices interleaved by A-atom (A = Al, Zn) mono- or bi-layers, had been extensively studied. The number of interleaved planes holding the Al atom and the M:B ratio determine the crystal structures of MAB compounds. Generally speaking, the proto-types of MAlB, $M_2AlB_2$, $M_3AlB_4$, $M_4AlB_6$, and $M_4AlB_4$ are called the 222, 212, 314, 416, and 414 MAB compounds, in that order [13,14]. The M–B blocks, that consist of the face-sharing trigonal prisms ($BM_6$), are the structural units of MAB phases. These blocks are separated by mono- or bi-layers of aluminum. The B atoms are located in the center of the $BM_6$ trigonal prisms. They are comparatively close to one another and form chains that are covalently bound [13].

The first MAB-type phase, MoAlB (reported as $Mo_7Al_6B_7$), was synthesized in 1942 by Halla and Thurry, was initially believed to crystallize in an orthorhombic structure with a *Pmmm* space group [15]. However, in 1966, Jeitschko et al. corrected this, showing that MoAlB crystallized in an orthorhombic structure but with a *Cmcm* space group, consisting of a CrB-like Mo–B sublattice interleaved with aluminum bilayers [16]. In the mid-1990s, further developments came, with the synthesis of solid-solution single crystals, specifically $(Mo_{1-x}W_x)AlB$ and $(Mo_{1-x}Cr_x)AlB$, expanding the known compositions and properties of MAB phases [17,18]. In 1973, Kuz'ma and Chaban discovered $Cr_2AlB_2$ [19]. The same authors also described a derivative structure of $Cr_2AlB_2$, which was composed of Al monolayers interleaved with a sublattice resembling $Cr_3B_4$ [19]. Ade and Hillebrecht found $Cr_4AlB_6$ in 2015; the metal boride sublattice of this compound replicates that of $Cr_2B_3$, or thick layers with 4 M atoms. This phase comprised the sequence $(CrB_2)_nCrAl$ (n = 1 – 3) of ternaries, together with $Cr_2AlB_2$ and $Cr_3AlB_4$ [14]. In 2016 Kota et al. [20] synthesized MoAlB in a dense, single-phase form by reactive hot pressing. Because of its structural resemblance to the MAX phases, this approach yielded a material with exceptional structural stability, which Kota et al. hypothesized would enable MoAlB to resist oxidation at high temperatures [20]. The many families of ternary TMBs result in an increase in structural diversity, some of which have been covered in previous publications [13,21]. Among them, the recently dubbed 'MAB phases' have received significant interest and include compounds with the following structure types: $M_2AlB_2$ (space group *Cmmm*), [22,23] MAlB (space group *Cmcm*), [16] $Cr_3AlB_4$ (space group *Immm*), [19] $Cr_4$- $AlB_6$ (space group *Cmmm*), [14], $Cr_4AlB_4$, [24], and $Ru_2ZnB_2$ (space group *$I4_1/amd$*) [25]. Comparable to the MAX phases, whereby the equivalent 2D materials were designated as MXenes following the etching of the "A" elements, the 2D transition metal borides derived from the MAB phases are referred to as MBenes [10]. Notably, 2D transition metal borides are finding a lot of use in magnetic and electrical fields [26–28]. According to recent estimates, there could be uses of 2D MBenes for Li- and Na-ion batteries, [29,30], electrocatalysis, [29] and magnetic [31] devices. Like 2D MXenes, [32,33] future developments in electrical, optoelectronic, and energy devices are expected to create a wide range of possible applications for MBenes.

Previously, a detailed description of the structural, elastic, electrical, optical, and bonding properties of the MoAlB material that was experimentally synthesized using the *ab-initio* approach [34] and several physical characteristics of NbAlB compound have been studied, such as, the structural, elastic, electronic, optical, bonding and thermal properties in the ground state using the density functional theory-based calculations [35]. However, to the best of our knowledge, no prior research has been done on any of the physical characteristics of VAlB and TaAlB compounds, including their structural, elastic, electronic, and chemical bonding, charge density distribution, Mulliken population analysis, thermomechanical properties, and energy-dependent optical constants. For example, there is currently no citation-worthy research available on the elastic properties of VAlB and TaAlB compounds, including the hardness, anisotropy in elastic moduli, machinability index, Poisson's ratio, and Kleinman parameter. Furthermore, no research has been done on the different thermo-mechanical characteristics of the selected compounds, such as the Debye temperature, thermal conductivity, melting temperature, sound velocities, and thermal expansion coefficient. Additionally, a theoretical knowledge of the bonding and optical properties of TaAlB and VAlB materials is currently lacking.

Investigating these undiscovered features is therefore of notable theoretical interest. We hope to address in-depth the structural, elastic, electronic, chemical bonding, charge density distribution, Mulliken population analysis, thermo-mechanical properties, and optoelectronic properties of VAlB and TaAlB compounds in this work, which is driven by the absence of fundamental data. We will also compare the results with those of previously studied MoAlB and NbAlB compounds [34,35]. Binary transition metal borides (TMBs) are well known for their hardness, electrical and thermal conductivities, high melting temperatures, inertness in a variety of corrosive circumstances and occasionally, for their magnetic characteristics [13]. By comparing MoAlB and NbAlB with VAlB and TaAlB, we may gain insight into their appealing physical features and find ways of their future engineering and optoelectronic applications.

This is how the remainder of the manuscript is organized: The computational technique is briefly described in Section 2. The examined physical properties and possible implications are presented and discussed in Section 3. Ultimately, we highlight the major results and make our work's conclusions in Section 4.

## 2. Computational scheme

The computational schema for the MoAlB and NbAlB that were previously reported elsewhere [34,35]. The density functional theory (DFT) was used to do first-principles calculations for the different characteristics of VAlB and TaAlB in the present investigation [36,37]. The quantum mechanical CAmbridge Serial Total Energy Package (CASTEP) code [38,39] was used. The exchange–correlation energy was evaluated using the generalized gradient approximation (GGA) within the Perdew-Burke-Ernzerhof (PBE) [40] scheme. This approach uses an ultra-soft Vanderbilt-type pseudopotential to simulate the interactions between ions and electrons [41] for V, Al and B atoms in VAlB compound and Ta, Al, B atoms in TaAlB compound. The basis sets of the valence electron states are set to be $3d^3$ $4s^2$ for V atoms, $3s^2$ $3p^1$ for Al atoms, and $2s^2$ $2p^1$ for B atoms in VAlB compound and $5d^3$ $6s^2$ for Ta atoms, $3s^2$ $3p^1$ for Al atoms, and $2s^2$ $2p^1$ for B atoms in TaAlB compound. The kinetic energy cut-off, which fixes the number of plane waves in the expansion, and the number of unique k-points that design the Brillouin zone (BZ) sampling are the two key factors that regulate the correctness of the computations. Step-by-step increase are made to the energy cut-off and the number of k-points until the estimated total energy converges within the necessary tolerance range. The plane-wave cut-off energy for both VAlB and TaAlB compounds were set at 400 eV. The Brillouin zone (BZ) sampling was carried out using the Monkhorst and Pack scheme [42] with a mesh size of 9 × 2 × 9 for both the compounds. Elastic constants, Mulliken bond population, charge density distribution, electronic band structure, density of states (DOS), and optical characteristics have been examined following the geometry optimization. When geometry was optimized for both VAlB and TaAlB compounds, the following tolerances were set, with finite basis set corrections: $< 5\times10^{-6}$ eV/atom for energy, $< 5\times10^{-4}$ Å for maximum lattice point displacement, < 0.01 eVÅ$^{-1}$ for maximum ionic force, and < 0.02 GPa for maximum stress [43]. Denser k-point meshes of sizes 35 × 8 × 37 and 37 × 8 × 38 were utilized for the VAlB and TaAlB, respectively, in the Fermi surface construction. All the properties studied refer to the ground state of the compounds.

Elastic constants for VAlB and TaAlB were computed using the stress-strain technique [44]. An orthorhombic crystal contains nine distinct second-order elastic coefficients ($C_{11}$, $C_{22}$, $C_{33}$, $C_{44}$, $C_{55}$, $C_{66}$, $C_{12}$, $C_{13}$, and $C_{23}$) when taking the crystal symmetry into account. Those separate elastic constants with the Voigte-Reuss-Hill (VRH) approximation, $C_{ij}$ enabled us to estimate all of the macroscopic (polycrystalline) elastic moduli, including the bulk modulus ($B$), shear modulus ($G$), and Young's modulus ($Y$) [45,46]. We investigate the bonding features in VAlB and TaAlB using the Mulliken population analysis (MPA) [47,48] and Hirshfeld population analysis (HPA) [47,49]. The charge density distributions in various crystallographic planes are also used for the same.

The compounds' frequency-dependent optical spectra may be obtained by applying the complex dielectric function, which is $\varepsilon(\omega) = \epsilon_1(\omega) + i\epsilon_2(\omega)$. Using the formula supported by CASTEP, the imaginary component of the dielectric function, $\epsilon_2(\omega)$, has been evaluated using the following expression:

$$\varepsilon_2(\omega) = \frac{2e^2\pi}{\Omega\varepsilon_0}\sum_{k,v,c}|\langle\psi_k^c|\hat{u}.\vec{r}|\psi_k^v\rangle|^2\,\delta(E_k^c - E_k^v - E) \quad (1)$$

In the above expression, $\Omega$ is the volume of the unit cell, $\varepsilon_0$ is the permittivity of the free space $\omega$ is the angular frequency (or equivalently energy) of the incident electromagnetic wave (photon), $\hat{u}$ is the unit vector defining the polarization direction of the incident electric field, $e$ is the electronic charge, $\psi_k^c$ and $\psi_k^v$ are the conduction and valence band wave functions at a given wave-vector $k$, respectively. Electronic band structure computations are used as inputs in this formula. Through the Kramers-Kronig transformation, the imaginary component of the dielectric function, $\epsilon_2(\omega)$, has been transformed into the real part, $\epsilon_1(\omega)$. After determining the dielectric function, one may use conventional formalism to determine all other optical constants, including refractive index $n(\omega)$, absorption coefficient $\alpha(\omega)$, energy loss-function $L(\omega)$, reflectivity $R(\omega)$, and optical conductivity $\sigma(\omega)$ [50,51].

Numerous thermophysical properties, including Debye temperature, melting temperature, thermal conductivity, thermal expansion coefficient, heat capacity, and others, are calculated using the computed elastic constants. By projecting the plane-wave states onto a linear combination of atomic orbital basis sets, the Mulliken bond population analysis was utilized to study the bonding characteristics of TaAlB and VAlB compounds [52,53]. The Mulliken bond population study was carried out using the Mulliken density operator, which may be expressed as follows on an atomic (or quasi-atomic) basis:

$$P_{\mu v}^{M}(g) = \sum_{g'} \sum_{g'} P_{\mu v'}(g') S_{v'v}(g - g') = L^{-1} \sum_{k} e^{-ikg} (P_k S_k)_{\mu v'} \qquad (2)$$

Moreover, the charge on atomic species A is expressed as follows:

$$Q_A = Z_A - \sum_{\mu \epsilon A} P_{\mu\mu}^{M}(0) \qquad (3)$$

where $Z_A$ refers to the charge of the atomic core.

## 3. Results and analysis

### 3.1 Structural properties

Many of a solid's characteristics, including optical qualities, electrical band structure, and elastic constants, are largely dependent on its crystal structure and symmetry. The arrangement and spacing between the atoms that make up a solid determine its physical and electrical characteristics. Compounds *M*AlB (*M* = V, Ta, Mo, Nb) crystallize in an orthorhombic unit cell with three distinct lattice constants (*a*, *b*, and *c*). Three kinds of atoms *M*, Al, and B, all occupy the 4c Wyckoff positions of space group $C_{mcm}$-$D_{2h}^{17}$ (No.63) [34,35]. This unit cell contains 12 atoms (4 *M*, 4 Al and 4 B) and has four formula units (Z = 4). The corresponding point group is *mmm*. A suitable description of the crystal structure of *M*AlB (*M* = V, Ta, Mo, Nb) compounds is given by the trigonal prismatic array of six *M* atoms surrounding each B atom, forming a nanolaminated structure as depicted in Fig. 1. It is discovered that two B and one Al atom are situated outside the trigonal prism's rectangular sides. The B atoms form B-B zigzag chains in the *c*-direction, while the prisms are packed so that all of the prism axes are parallel to the *b*-direction. The Al atoms form firmly wrinkled metal layers interleaved between the Mo double layers [54]. The optimized lattice parameters of VAlB and TaAlB compounds along with previously investigated MoAlB and NbAlB compounds are listed in Table 1. This table shows that the values of *a*, *b*, *c*, and *V* determined by the functional GGA (PBE) [40] deviate from the experimental values by not more than 0.004% and 0.017% for VAlB and TaAlB, respectively. These results imply an excellent agreement between the theoretical results and the experimental data.

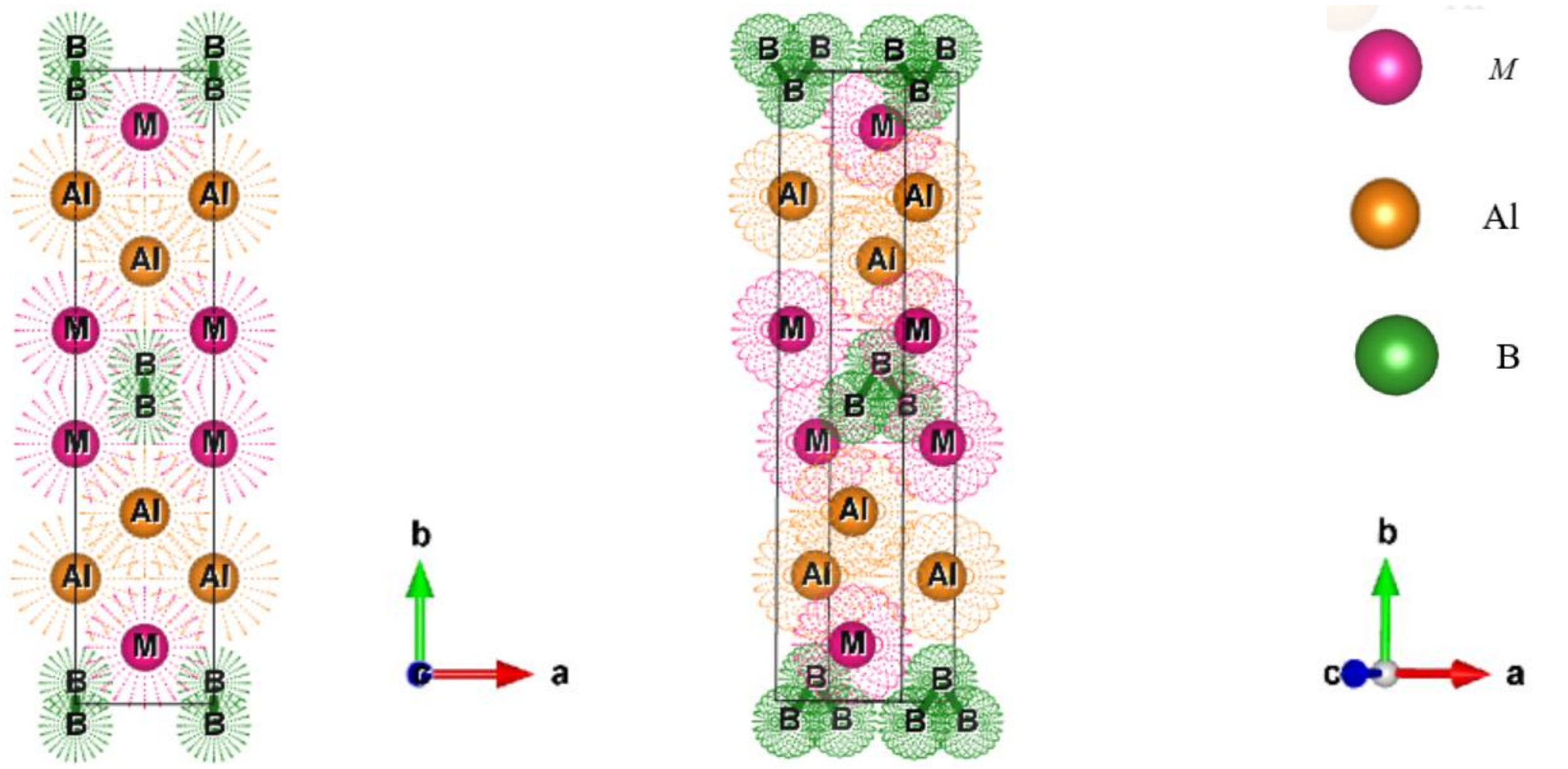

**Fig. 1.** Schematic crystal structure (2D and 3D) of *M*AlB (*M* = V, Ta, Mo, Nb) compounds. The crystallographic directions are shown.

Compared to the other compounds, TaAlB has a greater cell volume and lattice parameters. The primary cause of this difference is the compounds' different atomic radii. The order of atomic radii can be expressed as: Ta > Nb > Mo > V. The schematic crystal structure of *M*AlB (*M* = V, Ta, Mo, Nb) compounds are shown in Fig. 1. The newly investigated VAlB and TaAlB compounds in this paper are completely isostructural to the previously investigated MoAlB and NbAlB MAB compounds [34,35].

In order to investigate the chemical stability for VAlB and TaAlB compounds, the formation enthalpy has been computed, and the results are also presented in Table 1. The formation enthalpy ($\Delta H_f$) of VAlB and TaAlB compounds have been calculated using the following equation [55]:

$$\Delta H_f(MAlB) = \frac{E_{Total}(MAlB) - xE(M) - yE(Al) - zE(B)}{x+y+z} \tag{4}$$

Where, $E_{Total}(MAlB)$ is the total enthalpy of VAlB and TaAlB compounds; $E(M)$ is the total enthalpy of the V or Ta atom, $E(Al)$ is the total enthalpy of the Al atom and $E(B)$ is the total enthalpy of B atom in the solid state. We have $x$ = 1 for V and Ta atoms, $y$ = 1 for Al atoms and $z$ =1 for B atoms in the unit cell. The negative values of formation enthalpy imply that the two MAB phases studied in this work are chemically stable.

**Table 1:** Calculated and previously obtained experimental/theoretical lattice constants (*a*, *b*, and *c* all in Å), equilibrium volume (*V* in Å$^3$) and formation enthalpy ($\Delta H_f$ in eV/atom) of the *M*AlB (*M* = V, Ta, Mo, Nb) compounds.

| **Compound** | ***a*** | ***b*** | ***c*** | ***V*** | **Formation enthalpy** | **Ref.** |
|---|---|---|---|---|---|---|
| ValB | 3.09 | 14.23 | 3.01 | 132.43 | -6.91 | This work |
| | 3.09 | 14.20 | 3.01 | 131.91 | - | [10]$^{Expt.}$ |
| TaAlB | 3.35 | 14.69 | 3.13 | 154.03 | -7.11 | This work |
| | 3.33 | 14.64 | 3.11 | 151.45 | - | [10]$^{Expt.}$ |
| MoAlB | 3.21 | 14.04 | 3.10 | 140.28 | - | [34]$^{Theo.}$ |
| NbAlB | 3.34 | 14.71 | 3.12 | 153.11 | - | [35]$^{Theo.}$ |

3.2. Elastic and mechanical properties

The study of a material's elastic properties is important for technological advancement because it provides the information required to understand how the material will respond under different types of mechanical stress. When calculating the mechanical and dynamical characteristics of a crystal, the elastic constants are particularly relevant. A symmetric 6 × 6 matrix, known as the stiffness matrix or elastic matrix, holds a crystal's elastic constants. The elastic constants are closely connected to the mechanical characteristics of compounds, such as ductility, stiffness, brittleness, plasticity, and elastic anisotropy under static and time-varying stresses. The number of independent components in a matrix is determined by the symmetry limitations in the crystal class. Nine independent elastic constants ($C_{ij}$) are found in compounds having orthorhombic crystal structures: $C_{11}$, $C_{22}$, $C_{33}$, $C_{44}$, $C_{55}$, $C_{66}$, $C_{12}$, $C_{13}$, and $C_{23}$. The several additional elastic properties of our selected compounds are computed using these single crystal elastic constants. Table 2 displays the calculated values of the single crystal elastic constants for the materials we selected. In order to be mechanically stable, An orthorhombic crystal structure must satisfy the following Born-Huang [56] stability criteria:

$C_{11} > 0;\ C_{44} > 0;\ C_{55} > 0;\ C_{66} > 0;$

$C_{11}C_{22} > C_{12}^2;$

$$(C_{11}C_{22}C_{33} + 2C_{12}C_{13}C_{23}) > (C_{11}C_{23}^2 + C_{22}C_{13}^2 + C_{33}C_{12}^2) \quad (5)$$

Table 2 lists the nine independent elastic constants for *M*AlB (*M* = V, Ta, Mo, Nb) that were computed. It can be observed that all of the $C_{ij}$ values are positive, meeting the stability requirements stated previously. Consequently, the compounds *M*AlB (*M* = V, Ta, Mo, Nb) exhibit mechanical stability. Additionally, it can be observed that the ground state elastic constants found here correspond well with the MoAlB and NbAlB values that were previously computed [34,35].

The diagonal elastic constants $C_{11}$, $C_{22}$ and $C_{33}$, among the nine independent elastic constants, quantify the resistance to linear compression along the crystallographic *a*-, *b*-, and *c*-axes, respectively. For the compounds VAlB, MoAlB, and NbAlB, the estimated value of $C_{22}$ is less than $C_{11}$ and $C_{33}$, indicating that these ternary borides are more compressible along the *b*-axis than they are along the *a*- and *c*-axes. However, for the compound TaAlB, the calculated value of $C_{11}$ is smaller than $C_{22}$ and $C_{33}$. As a result, compared to the *b*- and *c*-axes, TaAlB is more compressible along the *a*-axis. The atomic bonding strengths in these MAB compounds are highest in the *c*-direction, as shown by the fact that $C_{33}$ > ($C_{11}$, $C_{22}$) for all compounds. $C_{12}$, $C_{13}$, and $C_{23}$ are the shear, off-diagonal components of the elastic constants. A uniaxial strain along the crystallographic *b*- and *c*-axes, respectively, is combined with a functional stress component in the crystallographic *a*-direction by the elastic tensors $C_{12}$ and $C_{13}$. The *M*AlB (*M* = V, Ta, Mo, and Nb) compounds should be able to resist shear along the crystallographic *b*- and *c*-axes when a significant force is applied along the crystallographic *a*-axis, according to the large values of these elastic components. For MoAlB and NbAlB compounds, $C_{23}$ has the lowest value among the three shear components. Together with a functional stress component along the crystallographic *b*-direction, it combines a uniaxial strain along the crystallographic *c*-direction. The observation of a significant magnitude difference between $C_{22}$ and $C_{33}$ is reflected in the lowest value of $C_{23}$. This implies that there is a significant difference between these two orientations in the strength of the intermolecular forces. The most important factor influencing a material's indentation hardness is $C_{44}$. The material's resistance to shear deformation in the (100) plane is shown by the large value of $C_{44}$, whereas the resistance to shear in the <011> and <110> directions is indicated by $C_{55}$ and $C_{66}$, respectively. It is discovered that MoAlB has a larger elastic constant $C_{44}$ than the other compounds considered. The ternary compound MoAlB should be able to resist shape changes and have greater levels of hardness than other compounds due to its substantially larger shear modulus and $C_{44}$ values.

**Table 2:** Single crystal elastic constants ($C_{ij}$ in GPa) of *M*AlB (*M* = V, Ta, Mo, Nb) compounds.

| Compound | $C_{11}$ | $C_{22}$ | $C_{33}$ | $C_{44}$ | $C_{55}$ | $C_{66}$ | $C_{12}$ | $C_{13}$ | $C_{23}$ | Ref. |
|---|---|---|---|---|---|---|---|---|---|---|
| ValB | 309.36 | 227.04 | 332.37 | 145.14 | 171.88 | 176.16 | 108.45 | 94.51 | 115.77 | This work |
| TaAlB | 283.43 | 290.53 | 345.28 | 156.08 | 176.52 | 177.81 | 114.53 | 163.60 | 116.49 | This work |
| MoAlB | 349.1 | 320.2 | 399.6 | 190.3 | 160.0 | 169.0 | 141.8 | 146.1 | 118.2 | [34] |
| NbAlB | 269.4 | 265.7 | 352.2 | 155.5 | 168.7 | 178.9 | 128.9 | 143.4 | 106.0 | [35] |

By using the Voigt-Reuss-Hill (VRH) approximations [57–59], the polycrystalline elastic moduli may be computed based on the determined elastic constants. The upper bound of the polycrystalline elastic moduli arises in the Voigt approximation when the discontinuous stress distribution inside the grains causes an imbalance in the real stresses among the grains [57]. On the contrary, the strain is distributed discontinuously inside the grains, whereas the stress is assumed to be continuous in the Reuss approximation. The polycrystalline elastic moduli's lower bound results from this discontinuity [58]. But the method known as Hill's approximation, which yields the closest approach to the practical polycrystalline elastic moduli. To evaluate the values of polycrystalline bulk modulus ($B$), shear modulus ($G$) and Young's modulus ($Y$), standard equations [60] are used. The Hill approximated values of bulk modulus ($B_H$) and shear modulus ($G_H$) (using the Voigt-Reuss-Hill (VRH) method), Young's modulus ($Y$) of $M$AlB ($M$ = V, Ta, Mo, Nb) compounds are estimated with the help of the following relations [32 –34]:

$$B_H = \frac{B_V + B_R}{2} \quad (6)$$

$$G_H = \frac{G_V + G_R}{2} \quad (7)$$

$$Y = \frac{9B_H G_H}{3B_H + G_H} \quad (8)$$

Table 3 contains a list of the obtained values and demonstrates how little the differences are between $G_V$ and $G_R$ and between $B_V$ and $B_R$. Hill states that the degree of elastic anisotropy of the solid should be proportional to the difference between these limiting values [59]. As a result, the ternary borides under investigation should have small anisotropy in elastic behavior. Bulk modulus represents a solid's capacity to withstand compression caused by uniform hydrostatic pressure. The shear modulus, on the other hand, measures the solid's capacity to withstand shape-changing external force. For all MAB phases, $B > G$, suggesting that the mechanical failure of VAlB, TaAlB, MoAlB, and NbAlB is governed by shearing stress rather than volume stress. The Young modulus quantifies the solid's resistance to length-changing tensile stress. The estimated Young's modulus values may be used to determine the stiffness and thermal shock resistance of the materials [61,62]. The values of $Y$ for the four compounds indicate that MoAlB is substantially stiffer than the other three compounds. Again, it is understood that the critical thermal shock resistance decreases inversely with the Young's modulus [62]. Thus, VAlB is projected to have higher thermal shock resistance than the other three materials.

The tetragonal shear modulus is an important measure for determining the dynamical stability of a crystalline solid [63,64]. It also delivers information on the rigidity of a crystal. Furthermore, it is related to slow transverse acoustic waves and plays an important role in the structural alterations of a system [65]. A positive value of $C_t$ ensures a crystal's dynamical stability, whereas a negative value of $C_t$ shows that the material is dynamically unstable. The tetragonal shear modulus ($C_t$) of $M$AlB ($M$ = V, Ta, Mo, and Nb) compounds may be determined using:

$$C_t = \frac{(C_{11} - C_{12})}{2} \quad (9)$$

The calculated value of $C_t$ for $M$AlB ($M$ = V, Ta, Mo, Nb) compounds are listed in Table 3. A positive $C_t$ implies that $M$AlB ($M$ = V, Ta, Mo, Nb) compounds are dynamically stable.

Materials can be classified as brittle or ductile based on indications such as Cauchy pressure ($C''$), Poisson's ratio ($\sigma$), and Pugh's ratio ($B/G$). The bonding of a molecule is reflected by the Cauchy pressure ($C''$) [66,67]. A substance of positive Cauchy pressure is often described as a ductile material, whereas brittleness is indicated by a negative value [68]. According to Pettifor [69], materials with a high positive value of $C''$ exhibit strong metallic non-directional bonding, whereas negative values correspond to angular bonding. The Cauchy pressure is zero for a bonding that can be characterized by simple paired potentials, such as the Lennard-Jones potential. The Cauchy pressure ($C''$) has been evaluated from the following formula:

$$C'' = (C_{23} - C_{44}) \quad (10)$$

Cauchy pressure as calculated for $M$AlB ($M$ = V, Ta, Mo, Nb) compounds are presented in Table 3. Cauchy pressure values below zero indicate brittleness and a high degree of angular bonding. The extremely negative Cauchy pressure of all $M$AlB ($M$ = V, Ta, Mo, Nb) compounds confirms their brittle nature. Additionally, high and negative values of $C''$ imply that there should be substantial angular (covalent) bonding between the atomic species in these MAB compounds.

The Poisson's ratio ($\sigma$) is a commonly used parameter to quantify ductility/brittleness, compressibility, and bonding force properties of a material. The Poisson's ratio also shows a compound's stability under shear. A smaller $\sigma$ value is associated with better shear stability. The numerical value of $\sigma$ is normally in the range of $-1.0 \leq \sigma \leq 0.50$. 0.26 is the crucial value of $\sigma$ at the polycrystalline material's border between brittleness and ductility [70,71]. If $\sigma$ is lesser (higher) than 0.26, a material is said to be brittle (ductile). The Poisson ratio may be used to anticipate the presence of central interatomic forces in materials, which are forces that act along the lines that unite pairs of atoms. In solids where central forces predominate in interatomic interactions, the Poisson's ratio often lies between 0.25 and 0.50; in other cases, the non-central force will predominate [72]. Furthermore, the value of $\sigma$ is around 0.33 for metallic compounds and 0.10 for pure covalent compounds. The Poisson’s ratio $(\sigma)$has been calculated using the formula [73,74] given below:

$$\sigma = \frac{(3B_H - 2G_H)}{2(3B_H + G_H)} \quad (11)$$

For $M$AlB ($M$ = V, Ta, Mo, Nb) compounds, the Poisson's ratio falls between these two typical values, suggesting that these borides have a combination of covalent and metallic bonding. The estimated values of $\sigma$ of $M$AlB ($M$ = V, Ta, Mo, Nb) compounds are listed in Table 3. Thus, we can predict that all the compounds are brittle, have a mixture of covalent and ionic bonds and the interatomic forces are non-central in nature. A low Poisson's ratio results from directed bonding, which raises a material's hardness by improving its shear modulus and limiting dislocation motion. Such dislocations tend to be minimized in materials with short covalent bonds, whereas they are allowed in materials with more delocalized bonds [75].

Pugh's ratio, or $B/G$, which represents a solid's bulk to shear modulus ratio, is used to determine whether a material is brittle or ductile [74,76,77]. At the boundary between brittle and ductile materials, the critical value of $B/G$ is around 1.75. Pugh's ratio demonstrates a relationship between ductility and value; a low value (<1.75) indicates brittleness in solids [77]. The brittle nature of these compounds is thus revealed by the Pugh's ratio for $M$AlB ($M$ = V, Ta, Mo, Nb) compounds ($B > G$). This suggests that the shearing stress, rather than the volume stress, should control the mechanical failure of $M$AlB ($M$ = V, Ta, Mo, Nb) compounds, which is consistent with the Cauchy pressure and Poisson's ratio.

**Table 3:** Polycrystalline bulk modulus ($B_V$, $B_R$, $B$ in GPa), shear moduli ($G_V$, $G_R$, $G$ in GPa), Young modulus ($Y$ in GPa), Pugh's ratio ($B/G$) and Poisson's ratio ($\sigma$), Cauchy pressure ($C''$ in GPa), tetragonal shear modulus, ($C_t$ in GPa) for the $M$AlB ($M$ = V, Ta, Mo, Nb) compounds in the ground state.

| Compound | $B_V$ | $B_R$ | $B$ | $G_V$ | $G_R$ | $G$ | $Y$ | $B/G$ | $\sigma$ | $C''$ | $C_t$ | Ref. |
|---|---|---|---|---|---|---|---|---|---|---|---|---|
| ValB | 167.36 | 164.09 | 165.72 | 135.30 | 119.76 | 127.53 | 304.49 | 1.30 | 0.19 | -29.37 | 100.46 | This work |
| TaAlB | 189.83 | 186.70 | 188.27 | 137.06 | 120.75 | 128.90 | 314.85 | 1.46 | 0.22 | -39.59 | 84.45 | This work |
| MoAlB | 209.0 | 207.5 | 208.0 | 148.2 | 139.1 | 144.6 | 351.9 | 1.44 | 0.22 | -72.1 | 70.25 | [34] |
| NbAlB | 182.7 | 179.9 | 181.3 | 134.6 | 116.6 | 125.6 | 306.1 | 1.44 | 0.22 | -49.5 | 103.65 | [35] |

We determined the Kleinman parameter ($\zeta$), representing a compound's stability under bending and stretching. The Kleinman parameter is dimensionless and typically has a value between 0 and 1. The Kleinman parameter qualitatively depicts the contribution of bond bending and stretching to the material's mechanical strength. Furthermore, the Kleinman parameter may be used to describe the relative shift of the cation and anion sub-lattice locations caused by volume conserving distortions in which the atomic positions are not fixed by crystal symmetry [78,79]. The Kleinman parameter ($\zeta$) can be calculated by the equation:

$$\zeta = \frac{C_{11}+8C_{12}}{7C_{11}+2C_{12}} \qquad (12)$$

Table 4 lists the computed Kleinman parameter values for VAlB and TaAlB compounds. Table 4 shows that the ζ value for ValB is less than 0.5, while TaAlB has a greater ζ value. This result led to the conclusion that mechanical strength in VAlB is mostly given by bond stretching, whereas mechanical strength in TaAlB is primarily provided by bond bending.

The machinability index ($\mu^M$) indicates how readily a material may be machined. This index may be used to assess the machinability of various solids and discover the best cutting settings for a certain material. In addition, this index may be utilized to analyze the plasticity and dry lubricating property of a material [60]. Plasticity increases with increasing machinability. Table 4 shows that the value of $\mu_M$ for all the $M$AlB ($M$ = V, Ta, Mo, Nb) compounds are almost similar but among them TaAlB shows the highest machinability index. The $\mu^M$ values of $M$AlB ($M$ = V, Ta, Mo, Nb) compounds (see Table 4) indicate high level of machinability, similar to several well-known MAX phase nanolaminates [80,81]. It is anticipated that all the MAB phases under study would have superior dry lubricating qualities, a reduced friction value, and lower feed forces. The machinability index was obtained using the given formula below:

$$\mu^M = \frac{B}{C_{44}} \qquad (13)$$

Higher bulk modulus alongside decreased shear resistance results in improved machinability and dry lubricity.

On heavy-duty devices, one of the main problems with surface hard coatings is the production of cracks, especially in metal and ceramic materials. An indicator of a material's resistance to surface crack development is its fracture toughness, or $F_T$. When designing engineering materials, this parameter is extremely important. The fracture toughness ($F_T$) may be computed using the following equation [35,82,83]:

$$F_T = V_0^{(1/6)} G(B/G)^{(1/2)} \qquad (14)$$

where, $V_0$ is the compound's volume per atom. In Table 4, the fracture toughness is also displayed. Table 4 shows that the NbAlB compound has a higher $F_T$ value than the other MAB compounds on the list. According to this finding, NbAlB will likely provide more resistance to the development of surface cracks than the other MAB compounds used in this study. The order of fracture toughness can be written as: NbAlB > TaAlB > MoAlB > VAlB. These MAB compounds have fracture toughness values that are significantly greater than those of several MAX phase borides. These compounds' high levels of hardness and fracture toughness make them appropriate for heavy-duty engineering applications [84].

The single crystal elastic constants or the pressure-dependent lattice parameter can be used to compute the bulk modulus of solids along the crystallographic axes. With the single crystal elastic constants, calculating the bulk modulus along three axes is simple. The relaxed bulk modulus and uniaxial bulk modulus along *a*-, *b*-, and *c*-axis, as well as anisotropies of the bulk modulus of VAlB and TaAlB compounds, have been evaluated using the following equations [85]:

$$B_{relax} = \frac{\Lambda}{(1+\alpha+\beta)^2} \qquad (15)$$

$$B_a = a\frac{dP}{da} = \frac{\Lambda}{1+\alpha+\beta} \qquad (16$$

$$B_b = b\frac{dP}{db} = \frac{B_a}{\alpha} \qquad (17)$$

$$B_c = c\frac{dP}{dc} = \frac{B_a}{\beta} \qquad (18)$$

$$A_{B_a} = \frac{B_a}{B_b} = \alpha \qquad (19)$$

$$A_{B_c} = \frac{B_c}{B_b} = \frac{\alpha}{\beta} \qquad (20)$$

where, $\Lambda = C_{11} + 2C_{12}\alpha + C_{22}\alpha^2 + 2C_{13}\beta + C_{33}\beta^2 + 2C_{33}\alpha\beta$

$$\alpha = \frac{\{(C_{11} - C_{12})(C_{33} - C_{13})\} - \{(C_{23} - C_{13})(C_{11} - C_{13})\}}{\{(C_{33} - C_{13})(C_{22} - C_{12})\} - \{(C_{13} - C_{23})(C_{12} - C_{23})\}}$$

$$\beta = \frac{\{(C_{22} - C_{12})(C_{11} - C_{13})\} - \{(C_{11} - C_{12})(C_{23} - C_{12})\}}{\{(C_{22} - C_{12})(C_{33} - C_{13})\} - \{(C_{12} - C_{23})(C_{13} - C_{23})\}}$$

where, $A_{B_a}$ and $A_{B_c}$ stand for bulk modulus anisotropies with respect to the *b*-axis along the *a*- and *c*-axes, respectively. Table 4 discloses the computed findings. The *c*-axis of VAlB and TaAlB compounds has a higher directional bulk modulus than the *a*- and *b*-axes. Thus, the compressibility along the *c*-axis is the lowest for both compounds. The bulk modulus of the TaAlB compound is bigger in the direction of the *b*- and *c*-axes than that of the VAlB compound, which is larger in the direction of the $a$-axis. In most cases, a material's uniaxial bulk modulus differs greatly from its isotropic bulk modulus. The reason for this is that, for a particular crystal density, the pressure under uniaxial strain is typically different from the pressure under hydrostatic stress at the same solid density [74]. Elastic isotropy is shown by a value of $A_{B_a}$ = $A_{B_c}$ = 1.0, while elastic anisotropy is indicated by any deviation from unity. These substances are therefore moderately anisotropic.

**Table 4:** The machinability index ($\mu^M$), Kleinman parameter ($\zeta$), fracture toughness ($F_T$ in MPa.m$^{1/2}$), bulk modulus ($B_{relax}$ in GPa), bulk modulus along *a*-, *b*- and *c*-axis ($B_a$, $B_b$, $B_c$ in GPa), and anisotropies of the bulk modulus along *a*- and *c*-axis ($A_{B_a}$and $A_{B_c}$) of *M*AlB (*M* = V, Ta, Mo, Nb) compounds.

| **Compound** | $\boldsymbol{\mu^M}$ | $\boldsymbol{\zeta}$ | $\boldsymbol{F_T}$ | $\boldsymbol{B_{relax}}$ | $\boldsymbol{B_a}$ | $\boldsymbol{B_b}$ | $\boldsymbol{B_c}$ | $\boldsymbol{A_{B_a}}$ | $\boldsymbol{A_{B_c}}$ | **Ref.** |
|---|---|---|---|---|---|---|---|---|---|---|
| ValB | 1.142 | 0.49 | 2.163 | 164.07 | 551.28 | 367.52 | 641.02 | 1.50 | 1.74 | This work |
| TaAlB | 1.206 | 0.54 | 2.382 | 186.71 | 519.03 | 459.32 | 798.51 | 1.13 | 1.74 | This work |
| MoAlB | 1.095 | - | 2.303 | - | - | - | - | - | - | [34] |
| NbAlB | 1.167 | - | 2.613 | - | - | - | - | - | - | [35] |

One of the most important mechanical qualities of materials that ensures quality in industrial applications is hardness, or the resistance to permanent deformation. Tribological coatings, heavy-duty goods, low-emission window glass coatings, optoelectronics, and microelectronics are among the application areas covered [86–88]. Understanding the link between hardness and other characteristics like a material's chemical stability, scratch resistance, and surface durability is crucial. Hard materials are used to make cutting tools and wear-resistant coatings. A relationship exists between a material's strength and hardness [89]. There are several theoretical frameworks for estimating the hardness as a function of Poisson's ratio ($\sigma$), Young's modulus ($Y$), bulk modulus ($B$), and shear modulus ($G$). Some of the prominent frameworks are given by N. Miao et al. [90,91], X. Chen et al. [83], Y. Tian et al. [92], and D. M. Teter [75], and E. Mazhnik et al. [93]. We have calculated the values of $(H_V)_{micro}$, $(H_V)_{macro}$, $(H_V)_{Tian}$, $(H_V)_{Teter}$, and $(H_V)_{Mazhnik}$ of VAlB and TaAlB compounds via the equations (21)-(25). Hardness values are disclosed in Table 5.

$$(H_V)_{Micro} = \frac{(1-2\sigma)Y}{6(1+\sigma)} \qquad (21)$$

$$(H_V)_{macro} = 2[\left(\frac{G}{B}\right)^2 G]^{0.585} - 3 \qquad (22)$$

$$(H_V)_{Tian} = 0.92(G/B)^{1.137}G^{0.708} \qquad (23)$$

$$(H_V)_{Teter} = 0.151G \qquad (24)$$

$$(H_V)_{Mazhnik} = \gamma_0 \chi(\sigma) Y \quad (25)$$

In Equation 25, $\chi(\sigma)$ is a function of Poisson's ratio and can be written as:

$$\chi(\sigma) = \frac{1 - 8.5\sigma + 19.5\sigma^2}{1 - 7.5\sigma + 12.2\sigma^2 + 19.6\sigma^3}$$

where $\gamma_0$ is a constant with a value of 0.096 and has no dimension.

As a result, hardness and shear as well as Young's modulus are correlated in addition to bulk modulus. The difference in the hardness values comes due to the parameters involved in the equations. However, $H_{micro}$ is observed to be higher than that of other hardness formulae obtained using Chen's formula and the formalism due to Mazhnik et al. [93] results in the lowest values of hardness. It is seen from Table 5 that the hardness of VAlB is greater than that of TaAlB, MoAlB and NbAlB. The order of the hardness values determined by various methods is as follows: VAlB > MoAlB > NbAlB > TaAlB. All these compounds appear to be fairly hard.

**Table 5:** Calculated hardness (in GPa) based on elastic moduli for *M*AlB (*M* = V, Ta, Mo, Nb) compounds.

| **Compound** | $(H_V)_{macro}$ | $(H_V)_{Tian}$ | $(H_V)_{Teter}$ | $(H_V)_{Micro}$ | $(H_V)_{Mazhnik}$ | **Ref.** |
|---|---|---|---|---|---|---|
| VAlB | 22.12 | 21.14 | 19.26 | 26.44 | 17.45 | This work |
| TaAlB | 18.86 | 18.65 | 19.46 | 24.06 | 15.11 | This work |
| MoAlB | 20.81 | - | - | - | - | [34] |
| NbAlB | 19.01 | - | - | - | - | [35] |

Anisotropy in mechanical characteristics is an important component that influences mechanical stability and structural stresses in a material under various forms of stress. For example, in engineering science, the formation and propagation of micro-cracks in materials, which is governed by mechanical property anisotropy, has a substantial impact on increasing a material's mechanical durability. In general, directed covalent bonding has a significant effect on crystal anisotropy, whereas metallic bonding helps to improve overall isotropy. The shear anisotropic factors can be used to determine the degree of anisotropy in atomic bonding across various crystal surfaces. The shear anisotropy of an orthorhombic crystal can be measured by three separate parameters [74,94]:

For {100} shear planes between the <011> and <010 > directions, the shear anisotropic factor is,

$$A_1 = \frac{4C_{44}}{C_{11}+C_{33}-2C_{13}} \quad (26)$$

between the <001> and <101> directions for the {010} shear plane is,

$$A_2 = \frac{4C_{55}}{C_{22}+C_{33}-2C_{23}} \quad (27)$$

and between the <110> and < 010> directions for the {001} shear planes is,

$$A_3 = \frac{4C_{66}}{C_{11}+C_{22}-2C_{12}} \quad (28)$$

Table 6 lists the computed shear anisotropic factors of $M$AlB ($M$ = V, Ta, Mo, and Nb) compounds. In the case of isotropy in shearing response, all three components must equal one. The departure from unity measures the degree of anisotropy. All of the compounds' calculated $A_1$, $A_2$ and $A_3$ values are more than unity. The estimated values predict that $M$AlB ($M$ = V, Ta, Mo, Nb) compounds are moderately anisotropic. The Zener anisotropy factor ($A$), universal anisotropy index ($A^U$, $d_E$), equivalent Zener anisotropy measure $A^{eq}$, anisotropy in shear $A_G$ and anisotropy in compressibility $A_B$ of solids with any symmetry can be estimated from the following standard formulas [95–98]:

$$A = \frac{2C_{44}}{C_{11}-C_{12}} \quad (29)$$

$$A^U = \frac{B_V}{B_R} + 5\frac{G_V}{G_R} - 6 \geq 0 \quad (30)$$

$$d_E = \sqrt{A^U + 6} \quad (31)$$

$$A^{eq} = \left(1 + \frac{5}{12}A^U\right) + \sqrt{\left(1 + \frac{5}{12}A^U\right)^2 - 1} \quad (32)$$

$$A_G = \frac{G_V - G_R}{G_V + G_R} \quad (33)$$

$$A_B = \frac{B_V - B_R}{B_V + B_R} \quad (34)$$

Tables 6 and 7 includes the results for these parameters. The universal elastic anisotropic index ($A^U$) is a popular metric for determining anisotropy as it applies to all crystal symmetries. The first anisotropy parameter, $A^U$, takes into consideration both shear and bulk contributions. Equation 30 shows that $G_V/G_R$ has a bigger impact on the anisotropy index $A^U$ compared to $B_V/B_R$. $A^U$ is 0 in locally isotropic single crystals, but deviation from zero indicates variable degrees of anisotropy in materials. The obtained values of $A^U$ for VAlB, TaAlB, MoAlB and NbAlB compounds are 0.67, 0.69, 0.33 and 0.79, respectively. Thus, according to this indication, the anisotropy in the NbAlB compound is greater than that of the other MAB compounds studied. For locally isotropic materials, $A^{eq}$ = 1.0. The computed values of $A^{eq}$ for VAlB and TaAlB are 2.08 and 2.09, respectively. These values predict that the examined compounds are somewhat anisotropic. Elastic isotropy ( $A_B$ = 0.0) and maximum anisotropy ($A_G$ = 100%) are represented by these numbers, respectively. The greater value of $A_G$ compared to $A_B$ (see Table 6) for $M$AlB ($M$ = V, Ta, Mo, Nb) compounds indicates more anisotropy in shear than compressibility. The universal log-Euclidean index can be defined using the log-Euclidean formula as [95,99]:

$$A^L = \sqrt{\left[ln\left(\frac{B_V}{B_R}\right)\right]^2 + 5\left[ln\left(\frac{C_{44}^V}{C_{44}^R}\right)\right]^2} \quad (35)$$

Here, $C_{44}^V$ and $C_{44}^R$ refer to the Voigt and Reuss values of $C_{44}$, respectively.

$$C_{44}^R = \frac{5}{3}\frac{C_{44}(C_{11} - C_{12})}{3(C_{11} - C_{12}) + 4C_{44}}$$

and,

$$C_{44}^V = C_{44}^R + \frac{3}{5}\frac{(C_{11} - C_{12} - 2C_{44})^2}{3(C_{11} - C_{12}) + 4C_{44}}$$

This index is accurately scaled for perfect isotropy and holds true for all crystallographic symmetries. $A^L$ is an absolute measure of anisotropy in crystalline materials, as it is less sparse than $A^U$ for strongly anisotropic crystals. The limitation of $A^U$ is that it merely describes the presence of anisotropy in a material, not the absolute level of it. $A^L$ is regarded a more accurate parameter for anisotropy analysis [95]. In a fully isotropic material, $A^L$ equals 0. Anisotropy increases with higher $A^L$ values. The projected $A^L$ values for VAlB and TaAlB are 0.21 and 0.55, respectively. Both are moderately anisotropic, although TaAlB exhibits greater anisotropy. Materials having greater (lower) $A^L$ typically have layered (or non-layered) structures [95,100]. The low value of $A^L$ indicates that the examined compounds have non-layered structural characteristics.

Solids' linear compressibility ($\beta_a$ and $\beta_c$) can usually be calculated using the formula [85]:

$$\beta_a = \frac{(C_{33}-C_{13})}{D} \quad (36)$$

$$\beta_c = \frac{(C_{11}+C_{12}-2C_{13})}{D} \quad (37)$$

Where, $D = (C_{11} + C_{12})C_{33} - 2(C_{13})^2$

The computed values for these parameters are shown in Table 7. For isotropic materials, the linear compressibility ratio ($\beta_c/\beta_a$) has a unit value. The departure of these components from unity represents the amount of anisotropic compressibility. The computed results imply that the linear compressibility of VAlB and TaAlB compounds is anisotropic. The estimated values for VAlB and TaAlB show that compressibility along the *a*-axis is larger than along the *c*-axis, which is consistent with the elastic constants.

**Table 6:** Shear anisotropy factors ($A_1$, $A_2$ and $A_3$), Zener anisotropy factor ($A$), anisotropy in shear ($A_G$) and anisotropy in compressibility ($A_B$) for *M*AlB (*M* = V, Ta, Mo, Nb) compounds.

| Compound | $A_1$ | $A_2$ | $A_3$ | $A$ | $A_G$ | $A_B$ | Ref. |
|---|---|---|---|---|---|---|---|
| ValB | 1.28 | 2.09 | 2.21 | 1.44 | 0.061 | 0.009 | This work |
| TaAlB | 2.07 | 1.75 | 2.06 | 1.85 | 0.063 | 0.008 | This work |
| MoAlB | 1.67 | 1.33 | 1.75 | - | 0.031 | 0.004 | [34] |
| NbAlB | 1.86 | 1.66 | 2.58 | - | 0.071 | 0.007 | [35] |

**Table 7:** Universal log-Euclidean index $(A^L)$, the universal anisotropy index $(A^U, d_E)$, equivalent Zener anisotropy measure $(A^{eq})$, linear compressibility ($\beta_a$ and $\beta_c$ in $TPa^{-1}$) and their ratio $(\beta_c/\beta_a)$ for $M$AlB ($M$ = V, Ta, Mo, Nb) compounds.

| Compound | $A^L$ | $A^U$ | $d_E$ | $A^{eq}$ | $\beta_a$ $(10^{-3})$ | $\beta_c$ $(10^{-3})$ | $\beta_c/\beta_a$ | Ref. |
|---|---|---|---|---|---|---|---|---|
| VAlB | 0.21 | 0.67 | 2.58 | 2.08 | 1.96 | 1.89 | 0.96 | This work |
| TaAlB | 0.55 | 0.69 | 2.59 | 2.09 | 2.16 | 0.84 | 0.38 | This work |
| MoAlB | - | 0.33 | - | - | - | - | - | [34] |
| NbAlB | - | 0.79 | - | - | - | - | - | [35] |

Elastic anisotropy of crystals may be effectively illustrated using the 2D and 3D graphical representations of materials' Young's modulus ($Y$), linear compressibility ($\beta$), shear modulus ($G$), and Poisson ratio ($\sigma$). CASTEP is used to compute the elastic stiffness matrices for VAlB and TaAlB compounds. This matrix is fed to the ELATE program [101] for direct visualization of anisotropy level in Young's modulus ($Y$), linear compressibility ($\beta$), shear modulus ($G$), and Poisson ratio ($\sigma$) of VAlB and TaAlB compounds. The spherical (3D) and uniform circular (2D) graphical representations are used as identifiers for the isotropic character of crystals. The degree of anisotropy increases with the variances from these forms. Figures 2 and 3 depict the direction dependency of $Y$, $\beta$, $G$, and $\sigma$ for TaAlB and VAlB compounds in 3D and 2D [in (xy)*ab*-, (xz)*ac*-, and (yz)*bc*-planes], respectively. The parameters' lowest and maximum points are shown, respectively, by the green and blue curves. Red curves indicate the negative values. The 3D plots of Young's modulus ($Y$), linear compressibility ($\beta$), shear modulus ($G$), and Poisson ratio ($\sigma$) show a larger deviation from spherical shape, suggesting significant elastic anisotropy. The 2D and 3D plots determine that in both compounds, anisotropy rises in the order $\beta < Y < G < \sigma$. Table 8 also includes a list of these parameters' upper and lower bounds as well as their ratios.

**Table 8:** The maximum and minimum values of Young's modulus ($Y$ in GPa), compressibility ($\beta$ in $TPa^{-1}$), shear modulus ($G$ in GPa), Poisson's ratio ($\sigma$) and their ratios for VAlB and TaAlB compounds.

| Compound | $Y$ | | $A_Y$ | $\beta$ | | $A_\beta$ | $G$ | | $A_G$ | $\sigma$ | |
|---|---|---|---|---|---|---|---|---|---|---|---|
| | $Y_{min}$ | $Y_{max}$ | | $\beta_{min}$ | $\beta_{max}$ | | $G_{min}$ | $G_{max}$ | | $\sigma_{min}$ | $\sigma_{max}$ |
| VAlB | 166.52 | 372.32 | 2.24 | 1.54 | 2.76 | 1.80 | 75.84 | 176.16 | 2.32 | -0.06 | 0.52 |
| TaAlB | 191.89 | 394.61 | 2.05 | 1.25 | 2.18 | 1.75 | 74.19 | 176.52 | 2.38 | -0.11 | 0.53 |

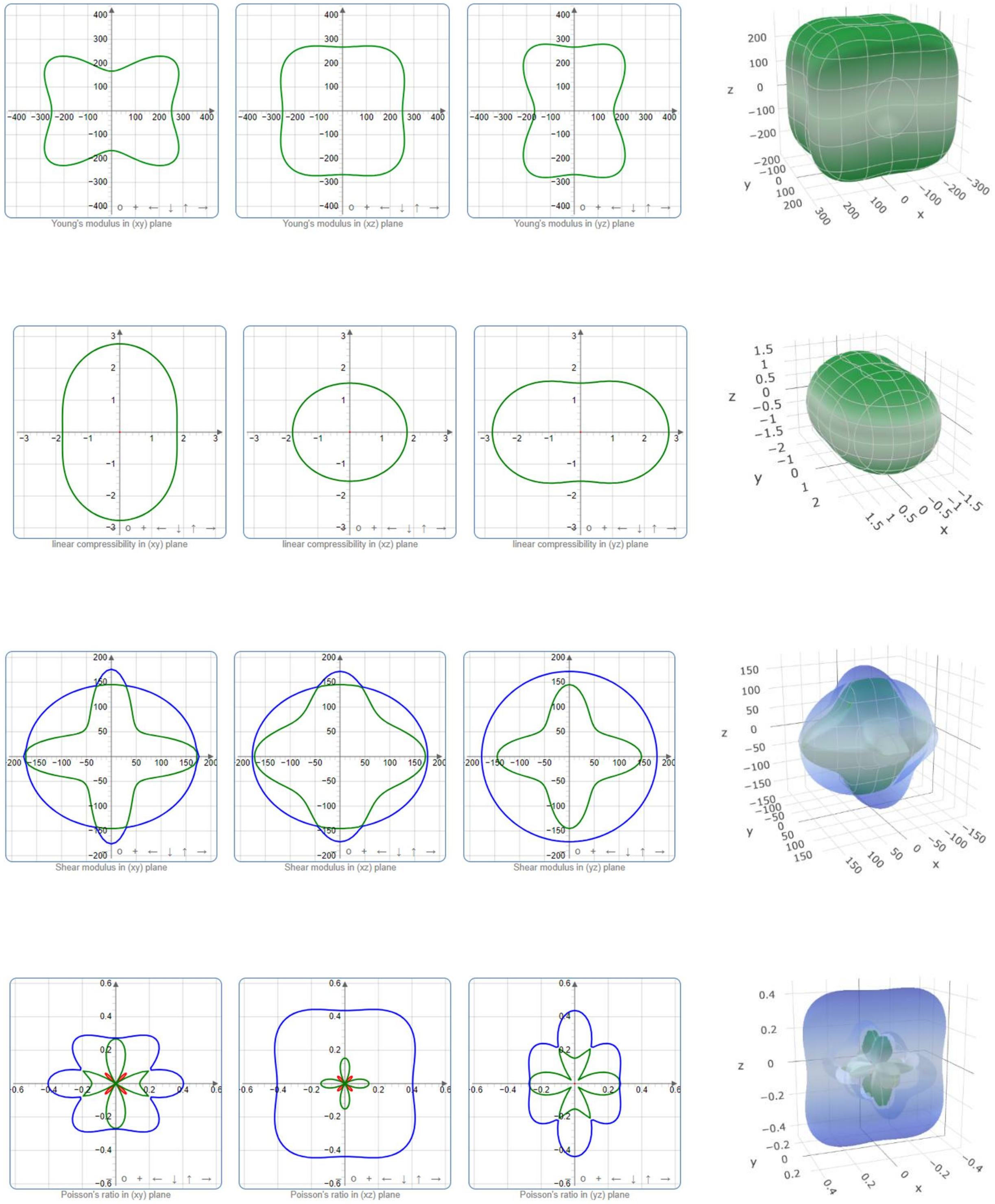

**Fig. 2.** Directional dependence of Young's modulus ($Y$), linear compressibility ($\beta$), shear modulus ($G$) and Poisson's ratio ($\sigma$) of VAlB compound.

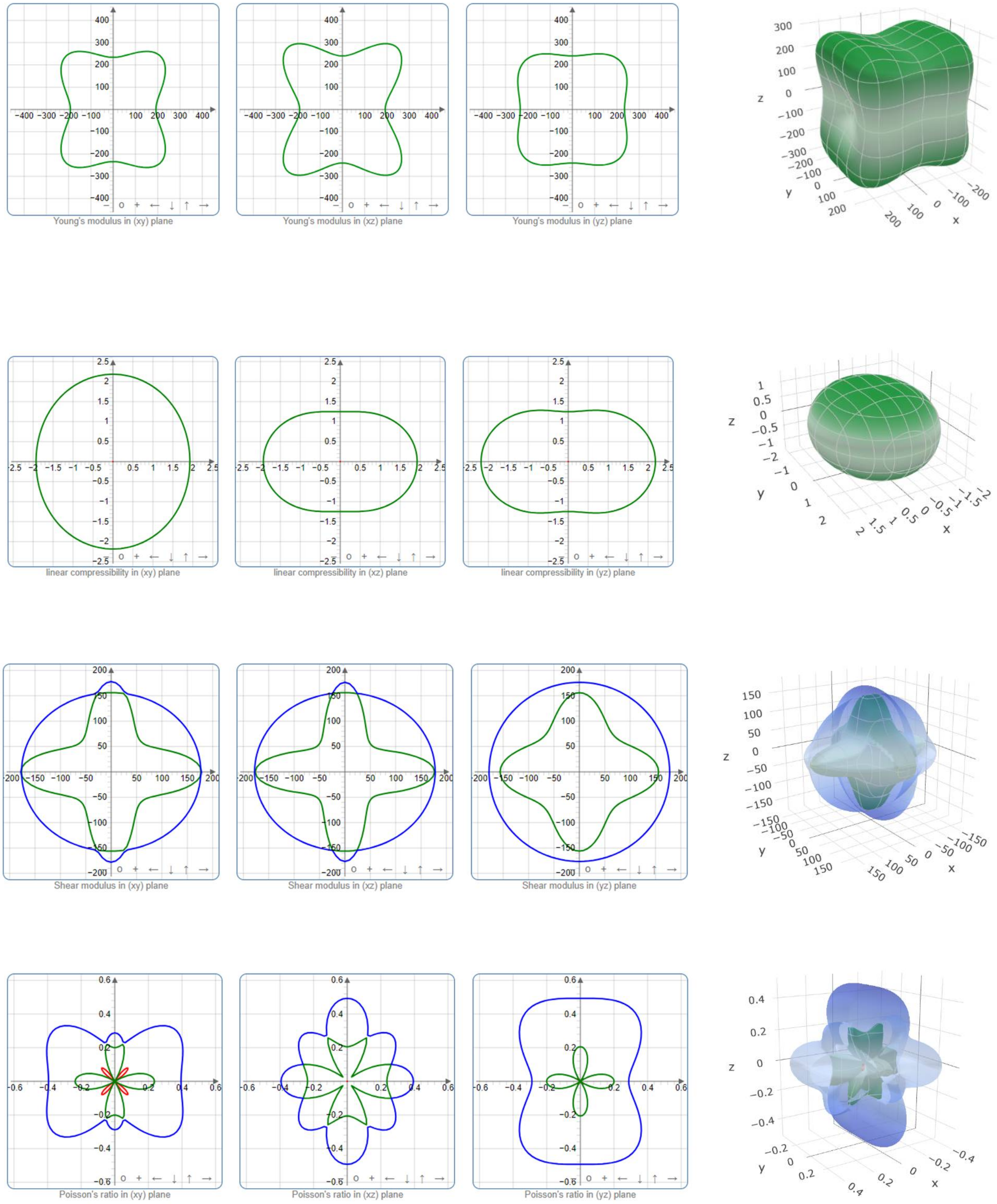

**Fig. 3.** Directional dependence of Young's modulus ($Y$), linear compressibility ($\beta$), shear modulus ($G$), and Poisson's ratio ($\sigma$) of TaAlB compound.

### 3.3. Mulliken and Hirshfeld population analysis

Atomic charges and charge transfer in compounds are two often utilized concepts in chemical bonding. Understanding bonding types (ionic, covalent, and metallic) in VAlB and TaAlB compounds, we have used both Mulliken population analysis (MPA) [48] and Hirshfeld population analysis (HPA) [49]. Table 9 shows the findings for the MAlB compounds (M = V, Ta, Mo, Nb). Mulliken charge examines how the electronic state changes in response to atomic displacements. This charge is also linked to the dipole moment, electric polarizability, and charge mobility in chemical processes, and other features of crystal structures. The charge spilling parameter measures the number of valence charges absent from the orbital projection. The smaller the parameter value, the better the model of electronic bonding. For the MAB phases, VAlB and TaAlB, V, Ta and Al contain positive Mulliken charge and B contains negative Mulliken charge; electrons are transferred from the V, Ta and Al atoms to the B atom, which is consistent with the previously investigated MoAlB and NbAlB compounds. The Mulliken charge value for V is +0.45e and Al is +0.14e in ValB compound, Ta is +0.48e and Al is +0.15e in TaAlB compound, Mo is +0.21e and Al is +0.28e in MoAlB compound and Nb is +0.37e and Al is +0.19e in NbAlB compound. This suggests that an ionic contribution to the bonding is present in these compounds. Therefore, in VAlB and TaAlB compounds, we may state that the V-B, Ta-B, Mo-B, Nb-B, and Al-B bonds have a partly ionic character. The degree of covalency and/or ionicity may be deduced from the values of effective valence charge (EVC), which is defined as the difference between the formal ionic charge and the Mulliken charge on the cation species. It establishes whether a link is ionic or covalent. A perfect ionic bond is formed when the effective valence charge is precisely equal to zero. Covalent bonding is indicated by an atom having an effective valence charge that is not zero. The degree of divergence from zero may be used to assess the covalent bonding level. In *M*AlB (*M* = V, Ta, Mo, Nb) compounds, the presence of both covalent and ionic bonds is implied by the non-zero values of EVC (see Table 9). The order of covalency level of the chemical bonds are: B-B > Al-Al > B-V > Al-V > B-Al in VAlB, B-B > Al-Ta > B-Ta > Al-Al > B-Al in TaAlB, B-B > Al-Al > Al-Mo > B-Mo > Al-B in MoAlB and B-B > Nb-Al > Al-Al > Al-B > Nb-B in NbAlB. The reverse of these orders should indicate the order of ionicity level of the chemical bonds in *M*AlB (*M* = V, Ta, Mo, Nb) compounds.

**Table 9:** Charge spilling parameter (%), orbital charges (electronic charge), atomic Mulliken charge (electronic charge), EVC (electronic charge), and Hirshfeld charge (electronic charge) of *M*AlB (*M* = V, Ta, Mo, Nb) compounds.

| Compound | Atoms | Charge spilling (%) | *s* | *p* | *d* | Total charge | Mulliken Charge | EVC | Hirshfeld Charge | Ref. |
|---|---|---|---|---|---|---|---|---|---|---|
| ValB | B | 0.31 | 0.98 | 2.61 | 0.00 | 3.59 | -0.59 | -2.41 | -0.14 | This work |
| | Al | | 0.91 | 1.95 | 0.00 | 2.86 | 0.14 | 2.86 | 0.06 | |
| | V | | 2.05 | 6.55 | 3.95 | 12.55 | 0.45 | 2.55 | 0.08 | |
| TaAlB | B | 0.78 | 1.07 | 2.57 | 0.00 | 3.64 | -0.64 | -2.36 | -0.14 | This work |
| | Al | | 0.99 | 1.86 | 0.00 | 2.85 | 0.15 | 2.85 | 0.01 | |
| | Ta | | 0.28 | 0.24 | 4.00 | 4.52 | 0.48 | 3.52 | 0.12 | |
| MoAlB | B | - | 0.95 | 2.53 | 0.00 | 3.49 | -0.49 | - | - | [34] |
| | Al | | 0.85 | 1.87 | 0.00 | 2.72 | 0.28 | 2.72 | - | |
| | Mo | | 2.11 | 6.47 | 5.21 | 13.79 | 0.21 | 5.79 | - | |
| NbAlB | B | - | 0.96 | 2.60 | 0.00 | 3.56 | -0.56 | - | - | [35] |
| | Al | | 0.89 | 1.93 | 0.00 | 2.81 | 0.19 | 2.81 | - | |
| | Nb | | 2.08 | 6.41 | 4.14 | 12.63 | 0.37 | 4.63 | - | |

3.4. Theoretical bond hardness

A material's resistance to deformation, abrasion, or indentation is measured by its hardness, a fundamental mechanical feature. It also characterizes a material's capacity to resist localized loads without suffering significant damage or plastic deformation. The relative hardness of a substance is also commonly used in industry. Hardness is an essential physical property of a material to understand its suitability, especially for usage as an abrasive and radiation-resistant element [102]. A solid is harder if it has a greater bond population, also known as electronic density, and shorter bond length. The overlap population of electrons between atoms indicates the bond's strength per unit volume. Hardness is classified into two types: intrinsic and extrinsic. In general, the hardness of a single crystal is considered intrinsic, while the hardness of polycrystalline materials is considered extrinsic. F. Gao [103] and H. Gou et al. [104] developed the Vickers hardness, a well-known theoretical approach for calculating intrinsic hardness. A solid with a greater bulk modulus and shear modulus often has a higher crystal stiffness and hardness. Although the bulk modulus and shear modulus provide some information on hardness, there is no direct link between hardness and bulk or shear modulus [105]. Bond hardness is a valuable intrinsic measure of a material and may be determined using the following equations [103,106]:

$$H_v^{\mu} = 740(P^{\mu} - P^{\mu\prime})\left(v_b^{\mu}\right)^{-5/3} \quad (38)$$

$$H_v = \left[\prod^{\mu}\left(H_v^{\mu}\right)^{n^{\mu}}\right]^{\frac{1}{\sum n^{\mu}}} \quad (39)$$

$$P^{\mu\prime} = \frac{n_{free}}{V} \quad (40)$$

$$n_{free} = \int_{E_P}^{E_F} N(E)\, dE \quad (41)$$

$$v_b^{\mu} = \frac{(d^{\mu})^3}{\sum_v [(d^{\mu})^3 N_b^{\mu}]} \quad (42)$$

where, $H_v^{\mu}$ is the bond hardness of μ-type bond, $H_v$ is the hardness of the compound, $P^{\mu}$ and $P^{\mu'}$ are the Mulliken bond overlap population and metallic population, $n_{free}$ is the number of free electrons, $V$ is the cell volume, $E_P$ and $E_F$ are the energy at the pseudogap and Fermi levels, respectively, $n^{\mu}$ denotes the total number of μ-type bonds, $v_b^{\mu}$ is the bond volume of μ-type bond, and $d^{\mu}$ is the bond length of μ-type bond. The constant 740 is a proportionality coefficient calculated from the hardness of the diamond. In this work, we estimated the metallicity ($f_m$) of VAlB, TaAlB and MoAlB compounds using the following formula [107]:

$$f_m = {P^{\mu'}}/{P^{\mu}} \quad (43)$$

Table 10 shows the computed number of free electrons, bond length, Mulliken and metallic population, total number of each kind of bond, metallicity, bond volume, bond hardness of each type of bond, and overall hardness of the compounds we studied. The Mulliken bond populations of a compound reflect the degree of overlap of the electron clouds that generate bonds between two atoms. For a completely ionic bond, the overlap population is zero, but any other positive number reflects the bond's covalency. The positive and negative values of a compound's Mulliken bond overlap populations may be used to determine the bonding and anti-bonding interactions between the relevant pairs of atoms, respectively [108]. Table 10 shows that all of the examined compounds exhibit positive bond populations for all bonds, implying that these atoms interact through bonding. Metallic populations for *M*AlB (*M* = V, Ta, Mo, Nb) compounds are extremely low (see Table 10). This result suggests the presence of a weak metallic link in these compounds. In these examined compounds, the B-Al bond has the highest metallicity, whereas the B-B bond has the lowest metallicity. Table 10 shows that the TaAlB compound has a greater total bond hardness value than the other compounds. Table 10 further shows that for *M*AlB (*M* = V, Ta, Mo, Nb) compounds, the B-B bond length is the shortest, with the highest bond overlap population. This means that almost the entire mechanical strength of *M*AlB (*M* = V, Ta, Mo, Nb) compounds is attributed to B-B atomic bonding. The general results in this section are consistent with the bond population study that was carried out earlier for MoAlB and NbAlB.

**Table 10:** Number of free electrons ($n_{free}$), bond length ($d^{\mu}$ in Å), bond number ($N^{\mu}$), Mulliken bond overlap population ($P^{\mu}$), metallic population ($P^{\mu'}$), metallicity ($f_m$), bond volume ($v_b^{\mu}$ in Å$^3$), bond hardness of μ-type bond ($H_v^{\mu}$ in GPa), and hardness ($H_v$ in GPa) of *M*AlB (*M* = V, Ta, Mo, Nb) compounds.

| **Compound** | $n_{free}$ | **Bond** | $d^{\mu}$ | $N^{\mu}$ | $P^{\mu}$ | $P^{\mu'}$ | $f_m$ | $v_b^{\mu}$ | $H_v^{\mu}$ | $H_v$ | **Ref.** |
|---|---|---|---|---|---|---|---|---|---|---|---|
| | | B-B | 1.751 | 2 | 1.47 | | 0.026 | 3.237 | 149.61 | | |
| | | B-Al | 2.296 | 4 | 0.08 | | 0.475 | 7.286 | 1.13 | | |
| ValB | 5.11 | B-V | 2.278 | 4 | 0.65 | 0.038 | 0.058 | 7.117 | 17.19 | 8.4 | This work |
| | | Al-Al | 2.700 | 2 | 0.82 | | 0.046 | 11.852 | 9.39 | | |
| | | Al-V | 2.664 | 4 | 0.56 | | 0.068 | 11.382 | 6.71 | | |
| TaAlB | 4.38 | B–B | 1.805 | 2 | 1.47 | 0.028 | 0.019 | 3.447 | 135.66 | 11.9 | This work |
| | | B–Al | 2.503 | 4 | 0.18 | | 0.156 | 9.192 | 2.79 | | |
| | | B–Ta | 2.455 | 4 | 0.96 | | 0.029 | 8.673 | 18.84 | | |
| | | Al–Al | 2.708 | 2 | 0.87 | | 0.032 | 11.641 | 10.42 | | |
| | | Al–Ta | 2.813 | 4 | 1.01 | | 0.028 | 13.048 | 10.05 | | |
| MoAlB | - | B-B | 1.809 | 2 | 1.38 | 0.004 | 0.003 | 3.570 | 122.01 | 11.6 | [34] |
| | | B-Al | 2.320 | 4 | 0.15 | | 0.032 | 7.531 | 3.71 | | |
| | | B-Mo | 2.369 | 4 | 0.66 | | 0.007 | 8.021 | 15.08 | | |
| | | Al-Al | 2.667 | 2 | 0.95 | | 0.005 | 11.437 | 12.04 | | |
| | | Al-Mo | 2.711 | 4 | 0.73 | | 0.006 | 12.013 | 8.51 | | |
| NbAlB | - | B-B | 1.770 | 2 | 1.52 | - | - | - | - | - | [35] |
| | | B-Al | 2.405 | 4 | 0.13 | | - | - | - | | |
| | | B-Nb | 2.368 | 4 | 0.56 | | - | - | - | | |
| | | Al-Al | 2.609 | 2 | 1.01 | | - | - | - | | |
| | | Al-Nb | 2.730 | 4 | 0.59 | | - | - | - | | |

3.5. Electronic charge density distribution

The electronic charge density distribution map displays precise information about the bonding type and amount of charge transfer between atoms in a molecule. It depicts the accumulation and depletion of electronic charges adjacent to various atomic species. The collection of charges between two atoms demonstrates covalent bonding between them. The presence of ionic bonds may be anticipated using a negative and positive charge balance at the atomic locations. To get a better understanding of chemical bonding between the ions of VAlB and TaAlB compounds, we examined the electronic charge density distributions depicted in Figs. 4 and 5 in the crystallographic planes (100), (010), and (001), respectively. The color scales show the intensity of total electronic density (e/Å$^3$). The colors blue and red signify high and low electronic densities, whereas red and blue represent high and low electronic densities in VAlB and TaAlB compounds, respectively. Fig. 4 shows that B has the largest charge density and V has the lowest charge density in all three planes of VAlB. Fig. 5 shows that in all three planes of TaAlB, B has the highest charge density and Ta has the lowest charge density. The electronic charge density distributions differ between the (100), (010), and (001) planes. Thus, the charge density distributions in VAlB and TaAlB compounds are anisotropic. Furthermore, the uneven charge distributions around the atoms are not circular. There are directional dependencies. This means that atomic bonding has both ionic and covalent components. The uniform charge backdrop also suggests the presence of metallic bonds in VAlB and TaAlB. The overall charge distribution in VAlB and TaAlB is qualitatively similar to that of MoAlB and NbAlB [34,35]. In all three planes of VAlB and TaAlB compounds, strong covalent bonding B-B exists between B-atoms, whereas weak covalent bonding B-Al exists between B- and Al-atoms, as shown in Figs. 4 and 5. Another covalent bond, Al-Al, occurs between Al-atoms, resulting in the formation of the Al layer in the molecule. The increase of charge also shows a covalent link between B-*M* and Al-*M*.

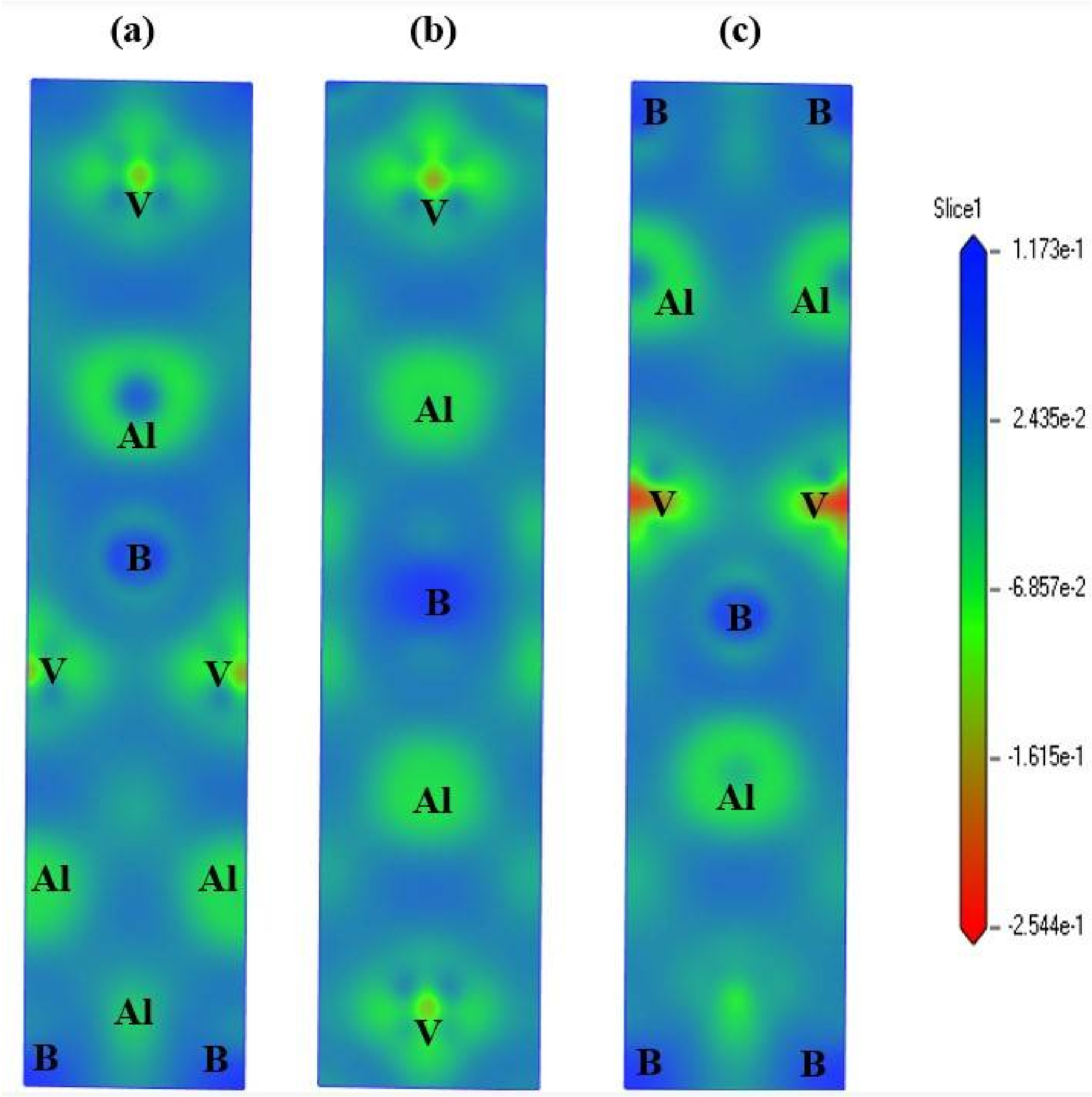

**Fig. 4.** The electronic charge density map in the (a) (100), (b) (010) and (c) (111) planes of VAlB compound. The charge density scale is shown on the right.

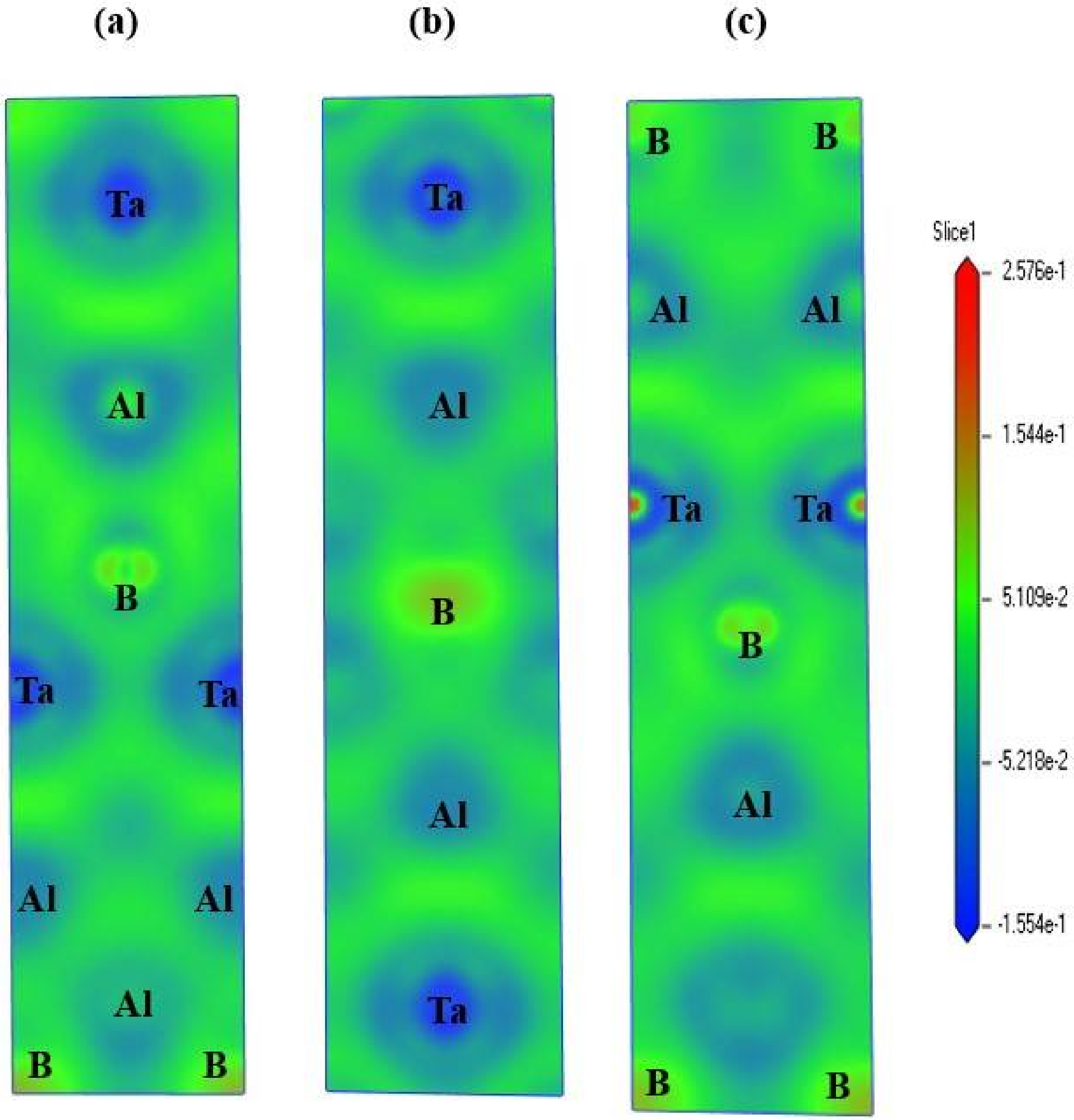

**Fig. 5.** The electronic charge density map in the (a) (100), (b) (010) and (c) (111) planes of TaAlB compound. The charge density scale is shown on the right.

### 3.6. Electronic properties
### 3.6.1. Electronic band structure

The study electronic band structure has high significance for explaining numerous microscopic and macroscopic phenomena such as chemical bonding, electronic transport, superconductivity, optical response, and magnetic order. Electronic characteristics are dominated by the bands close to the Fermi level. Modeling nanostructures and electrical devices requires good band topologies and suitable effective charge carrier masses. Studying the electronic band structure of materials is crucial in catalyst design [109], accelerated "battery materials" discovery [110], magnetism, and superconducting materials development [111].

We estimated the electronic band structures for VAlB and TaAlB along high symmetry directions (Γ - Z - T - Y - S - X - U - R) of the first BZ in the energy range of −5 to +5 eV as a function of energy ($E$-$E_F$), as shown in Fig. 6. The Fermi level $E_F$ is illustrated as a horizontal dotted line in Fig. 6. TaAlB has 79 bands, while VAlB has 93. The band structure of $M$AlB ($M$ = V, Ta, Mo, Nb) compounds demonstrates their metallic character, since there is a considerable overlap of conduction and valence bands with variable degrees of dispersion across the Fermi level [34,35]. There are six bands for VAlB and TaAlB compounds crossing the Fermi level which are shown in different colors with their corresponding band numbers. For VAlB compound bands 35, 36, 37, 38, 39 and 40 cross the Fermi level whereas, bands 19, 20, 21, 22, 23 and 24 cross the Fermi level for the TaAlB compound (see Fig. 6). Highly dispersive bands results in a low charge carrier effective mass [112–114] and high charge mobility. Each material has electron-like and hole-like properties in the BZ that extend in different directions. The band structure calculations also help us understand the shape of the underlying Fermi surfaces.

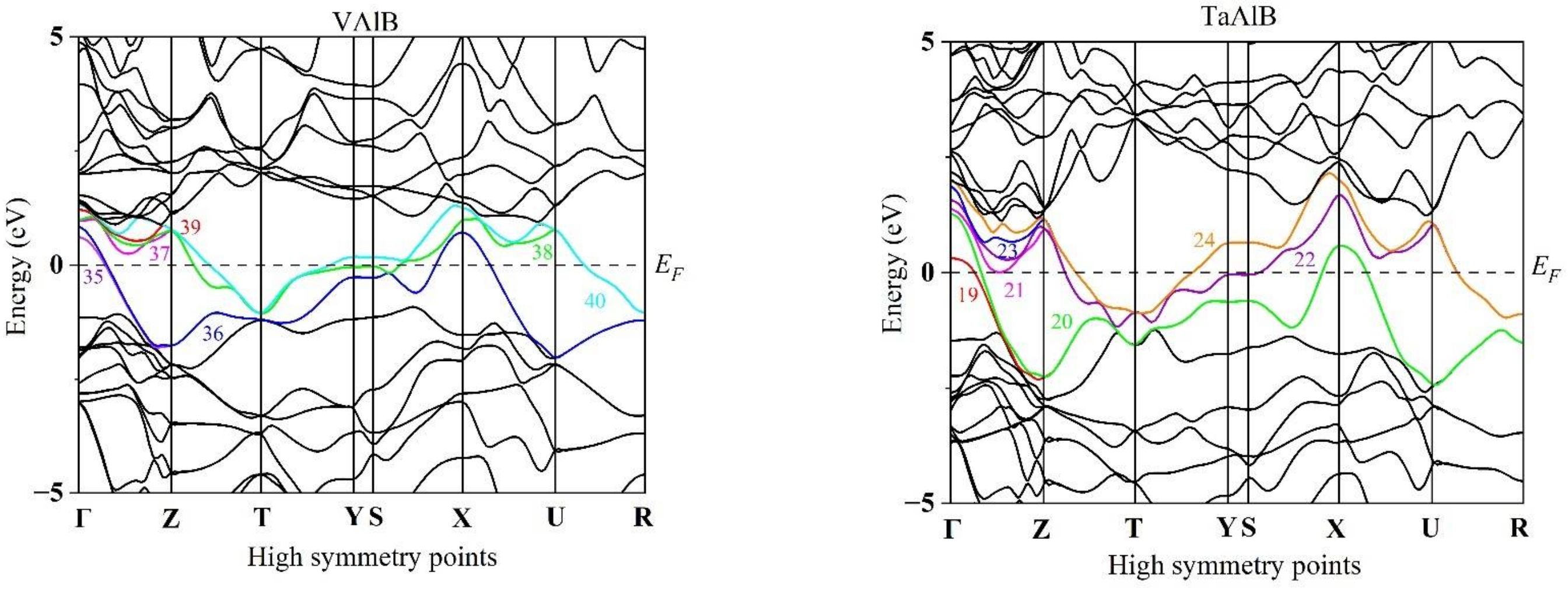

**Fig. 6.** The calculated electronic band structures of the VAlB and TaAlB compounds along several high symmetry directions of the Brillouin zone in the ground state.

It is evident from Fig. 6 that bands running along the $c$-direction are quite flat. These bands consist of heavy charge carriers with high degree of localization. On the other hand, charge carriers within the $ab$-plane are light and are expected to be highly mobile. Therefore, there is electronic anisotropy in the MAB phases considered here.

3.6.2. Electronic energy density of states (DOS)

The density of states (DOS) is the number of electronic states per unit energy per unit volume that are accessible for occupy. A high DOS for a given energy level indicates that there are several states accessible for occupancy. In this work, we calculated the total and partial density of states (TDOS and PDOS) of VAlB and TaAlB compounds to acquire a better knowledge of their electronic properties. Fig. 7 depicts the results of the TDOS and PDOS calculations performed on our chosen materials' GGA (PBE). The previous investigations examined the electronic energy density of the states of MoAlB and NbAlB [34,35].

The Fermi level ($E_F$) is shown by the vertical broken line at zero energy. It is evident from Fig. 7 that the TDOS values are finite at the Fermi level for both compounds. The non zero values of TDOS at the Fermi level for VAlB and TaAlB compounds is the evidence that like MoAlB and NbAlB [34,35], these two compounds (VAlB and TaAlB) should also exhibit metallic behavior. The value of TDOS at the $E_F$ for VAlB and TaAlB compounds are 6.97 states/eV-unit cell and 3.75 states/eV-unit cell, respectively. The corresponding value for the MoAlB and NbAlB compounds are 2.90 states/eV-unit cell and 5.45 states/eV-unit cell, respectively [34,35]. Consequently, compared to other studied MAB compounds, the TDOS of VAlB compound is noticeably higher. Consequently, we anticipate that VAlB's electrical conductivity will be significantly greater than that of the other MAB compounds under investigation. The TDOS at the Fermi level is primarily formed by the contribution of the V 3*d* electronic states and the Ta 5*d* electronic states in VAlB and TaAlB compounds, respectively. The Al 3*p* and B 2*p* orbitals also contribute to the properties of VAlB and TaAlB compounds. There is strong hybridization between the V 3*d*, Al 3*p*, and B 2*p* electronic states of the VAlB compound and the 5*d*, Al 3*p*, and B 2*p* electronic states hybridized for the TaAlB compound close to the Fermi level. Such high hybridization results in the development of covalent bonds between the orbitals involved [115]. The conduction band of VAlB and TaAlB compounds are dominated by the V 3*d* electronic states Ta 5*d* electronic states. The bonding peak is the closest peak at the negative energy below the Fermi level in the TDOS, while the anti-bonding peak is the closest peak at the positive energy. The pseudo-gap, or energy gap between these peaks, is a sign of the electrical stability involved [60,116,117]. A pseudogap exists for the previously studied MAB compound MoAlB close to the Fermi level [34] which helps to stabilize the structure of MoAlB and suppresses TDOS while acting as a boundary between bonding and anti-bonding electronic states. In contrast, VAlB, TaAlB, and NbAlB compounds do not have a pseudogap [35].

Coulomb interactions in metallic systems are investigated. The Coulomb pseudopotential ($\mu^*$) indicates the amount of the screening and effective Coulomb interaction. A material's electron-electron interaction parameter, known as the repulsive Coulomb pseudopotential, may be determined using the electronic density of states at the Fermi energy. The repulsive Coulomb pseudopotential, may be determined using the electronic density of states at the Fermi energy as follows [118]:

$$\mu^* = \frac{0.26\, N(E_F)}{1 + N(E_F)} \qquad (44)$$

where, $N(E_F)$ is the total density of states at the Fermi level of the compound. The value of $N(E_F)$ for VAlB and TaAlB compounds are 6.97 states/eV-unit cell and 3.75 states/eV-unit cell, respectively. The electron-electron interaction parameter VAlB and TaAlB compounds are therefore found to be 0.16 and 0.13, respectively, implying stronger electronic correlations in VAlB than TaAlB. The relatively high value of this parameter for TaAlB results from the high TDOS at the Fermi level, which is mostly due to the 3*d* electrons of V atoms and the 5*d* electrons of Ta atoms. The repulsive Coulomb pseudopotential is responsible for decreasing the transition temperature, $T_c$, of superconducting materials [118,119].

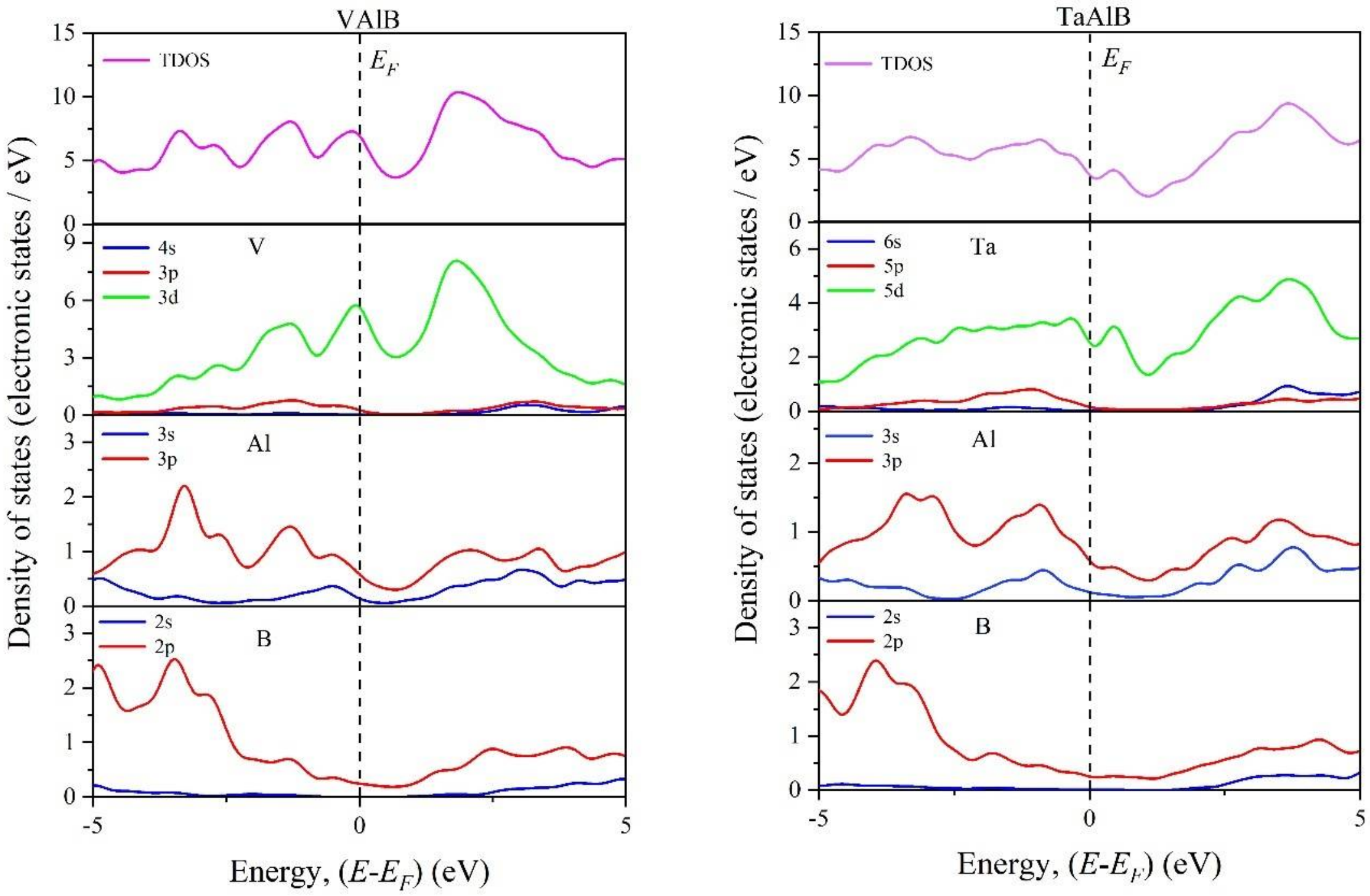

**Fig. 7.** The total and partial density of states (TDOS and PDOS, respectively) plots of VAlB and TaAlB as a function of energy.

3.6.3. Fermi surface

The Fermi surface should be explored in order to understand the behavior of electrons in a metallic material at low temperature. Several features, including electrical, optical, thermal, and magnetic, are highly dependent on Fermi surface topology. Superconducting state involves electrons near the Fermi sheets [120]. We have constructed the Fermi surface of VAlB and TaAlB compounds from the electronic band structures, as shown in Fig. 8. Fermi surfaces are constructed from the bands 35, 36, 37, 38, 39 and 40 for VAlB and 19, 20, 21, 22, 23 and 24 for TaAlB which cross the Fermi level. ValB and TaAlB compounds exhibit both electron and hole-like sheets in their Fermi surface topology [34]. First two sheets are centered along Γ–X direction in VAlB and TaAlB. The first sheet looks like the distorted or deformed cylindrical shape for VAlB and TaAlB. The second sheet appears to be more polyhedral or faceted compared to first deformed cylindrical surface for both compounds. The third, fourth, fifth and sixth sheet are almost similar and appears as curved, sheet-like structures extending through multiple directions in the Brillouin zone, potentially representing an open structure (electrons or holes are not confined to isolated areas). The surfaces are highly anisotropic, with varying shapes and orientations in different regions. This indicates complex electronic behavior, where the conduction properties are directionally dependent. The combined sheets produce a complicated topology of the Fermi surface. The low-dispersive V 3*d*-like bands and Ta 5*d*-like bands contribute to the Fermi surface of VAlB and TaAlB compounds, respectively. It also determines the conductivity of compounds. The electronic band structure and Fermi surface topology show good agreement with those observed in *M*AlB (*M* = Mo, Nb) [34,35].

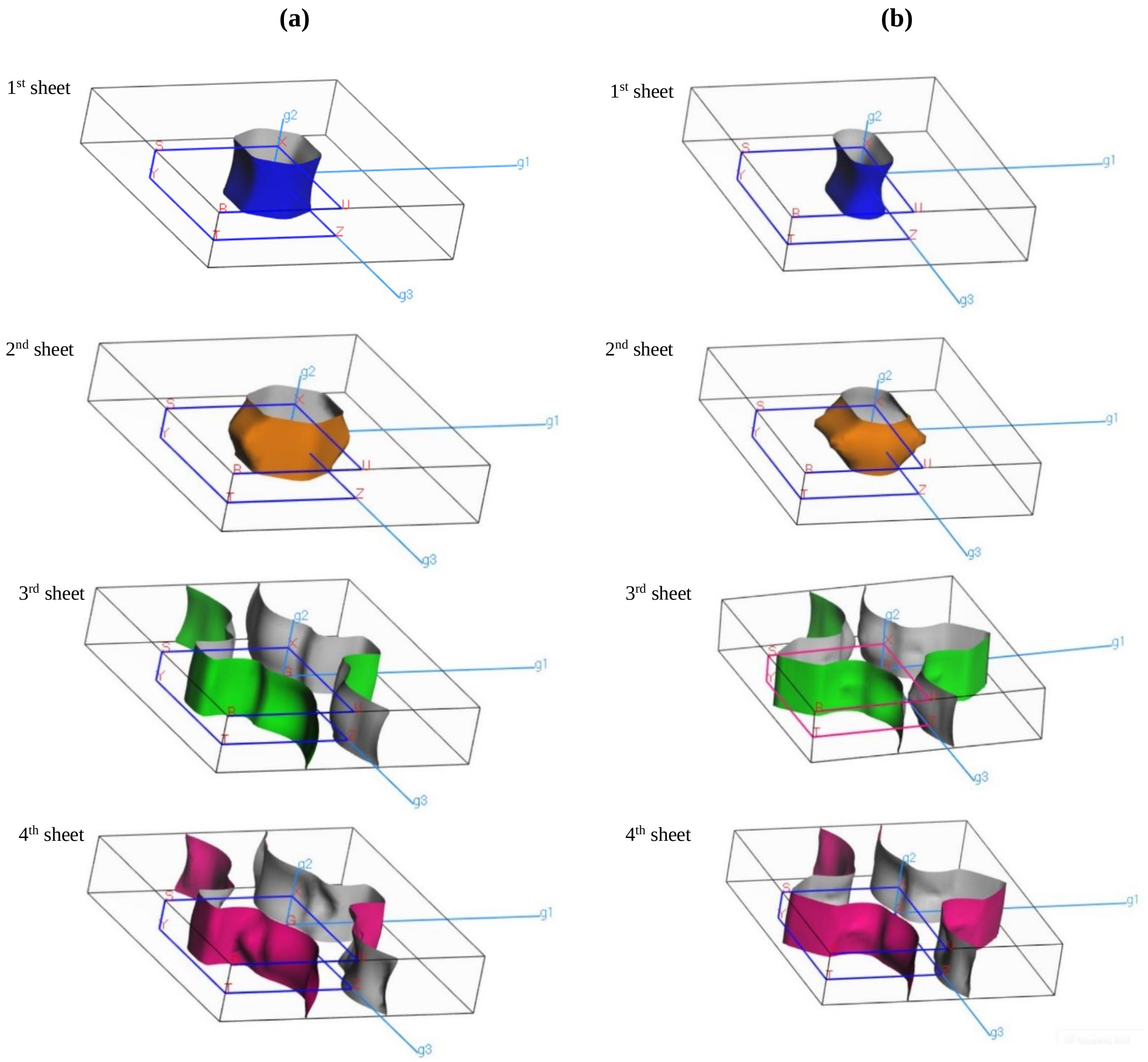

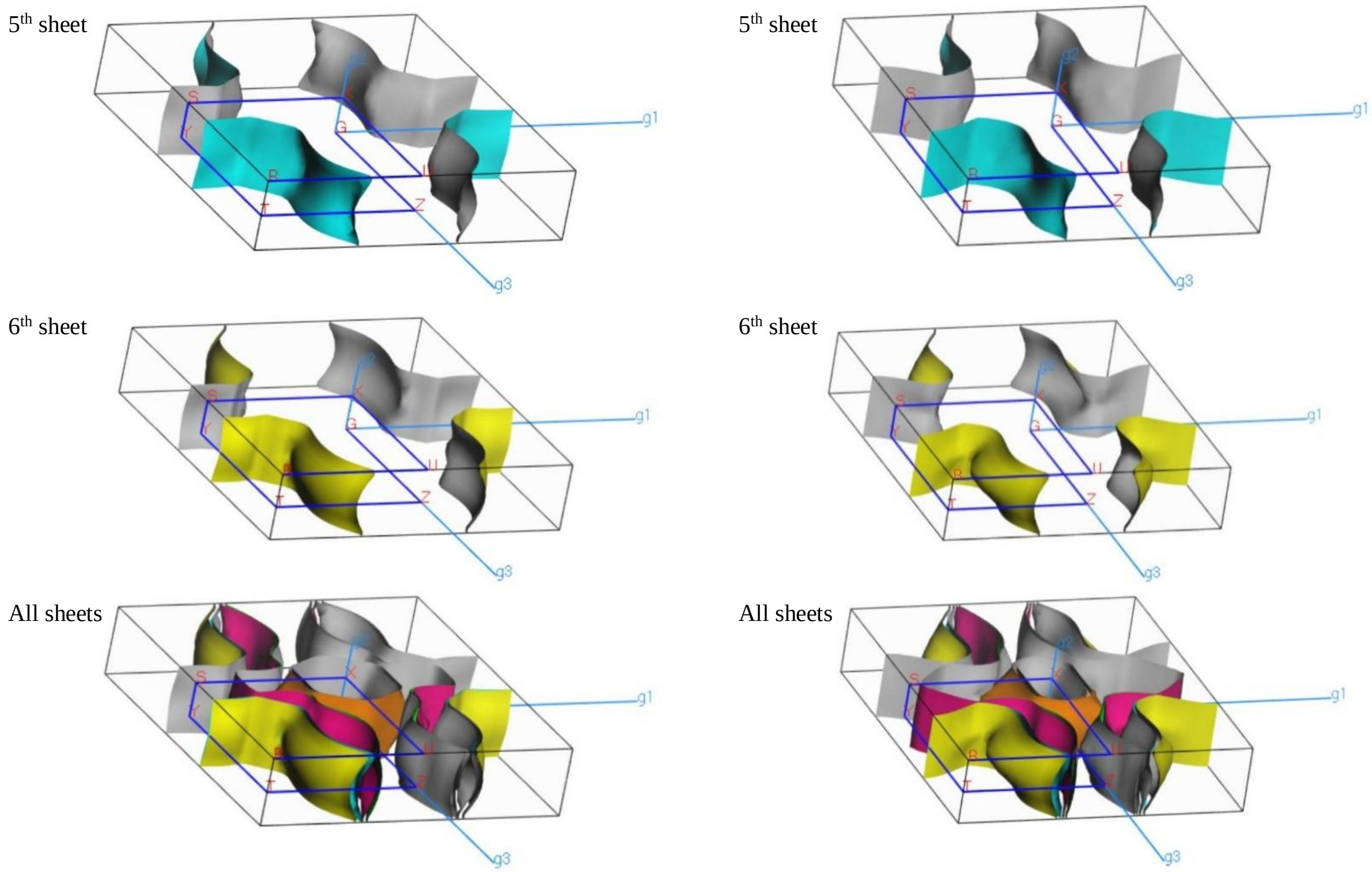

**Fig. 8.** Fermi surface topology of (a) VAlB compound and (b) TaAlB compound with individual sheets.

3.7. Phonon dynamics

The computation of phonon dispersion spectra (PDS) and phonon density of states (PHDOS) in crystalline materials has become an essential research aspect. It is informative to check as it provides a lot of relevant information about the material related dynamic lattice stability/instability, phase transition and vibrational contribution to heat conduction, thermal expansion, superconducting $T_c$, Helmholtz free energy, and heat capacity [121–123]. The functional GGA(PBE) is used to compute the phonon dispersion spectra along the Γ - Z - T - Y - S - X - U - R paths in the Brillouin zone (BZ) and the phonon density of states (DOS) of VAlB and TaAlB compounds. This computation has been done using a finite displacement technique based on density functional perturbation theory (DFPT) [124,125]. The phonon spectrum, as well as the electron-phonon coupling constant, have a significant impact on superconductivity below critical temperature. The dynamical stability of a material is an important criterion for its applicability under time variable mechanical stress. Fig. 9 depicts the predicted phonon dispersion curves and phonon DOS along high symmetry directions within the BZ of VAlB and TaAlB compounds. The presence of negative phonon frequency guarantees dynamic instability. There is no negative frequency branch found in the dispersion curves (see Fig. 9). This means that the VAlB and TaAlB crystals are dynamically stable in the ground state.

The materials' orthorhombic structure has a unit cell of 12 atoms, resulting in 36 normal lattice vibration modes, including 3 acoustic (pink) and 33 optical (blue) modes. The acoustic modes have one longitudinal and two transverse acoustic branches. Acoustic phonons are created by atoms in a lattice vibrating coherently about their equilibrium position. The optical phonon, on the other hand, is caused by out-of-phase oscillations of atoms in the lattice when one travels to the left and other moves to the right. Acoustic phonons have a role in sound propagation in crystals and are associated with crystal rigidity. The 3 acoustic modes occur at low frequencies, whereas the 33 optical modes occur at high frequencies, with some overlap for VAlB and TaAlB (see Fig. 9). Acoustic modes with low frequencies are often formed by the vibration of a heavy atom, whereas optical modes with high frequencies are produced by the vibration of a light atom. VAlB and TaAlB compounds do not have a phonon gap. In VAlB compound, the optical branches extend to significantly higher energies than in TaAlB compound. The highest vibrational frequency for VAlB and TaAlB compounds occurs around the Γ-point of the BZ with 25.49 THz and 22.25 THz, respectively. This occurs since the V atom has low atomic mass than Ta atom. To get a better understanding of lattice dynamics, we have shown total and atomic partial PHDOS of VAlB and TaAlB compounds in Fig. 9, coupled with PDC to identify the bands and their related densities of states. The PHDOS shows strong peaks at 9.07 THz and 4.45 THz for VAlB and TaAlB, respectively. The flatness of the bands causes peaks in the PHDOS, whereas highly dispersive bands reduce the heights of peaks in the overall PHDOS. The top and lower branches of phonon dispersion curves coincide, hence there is no phononic band gap between acoustic and optical modes in the PHDOS of VAlB and TaAlB compounds.

**(a)** **(b)**

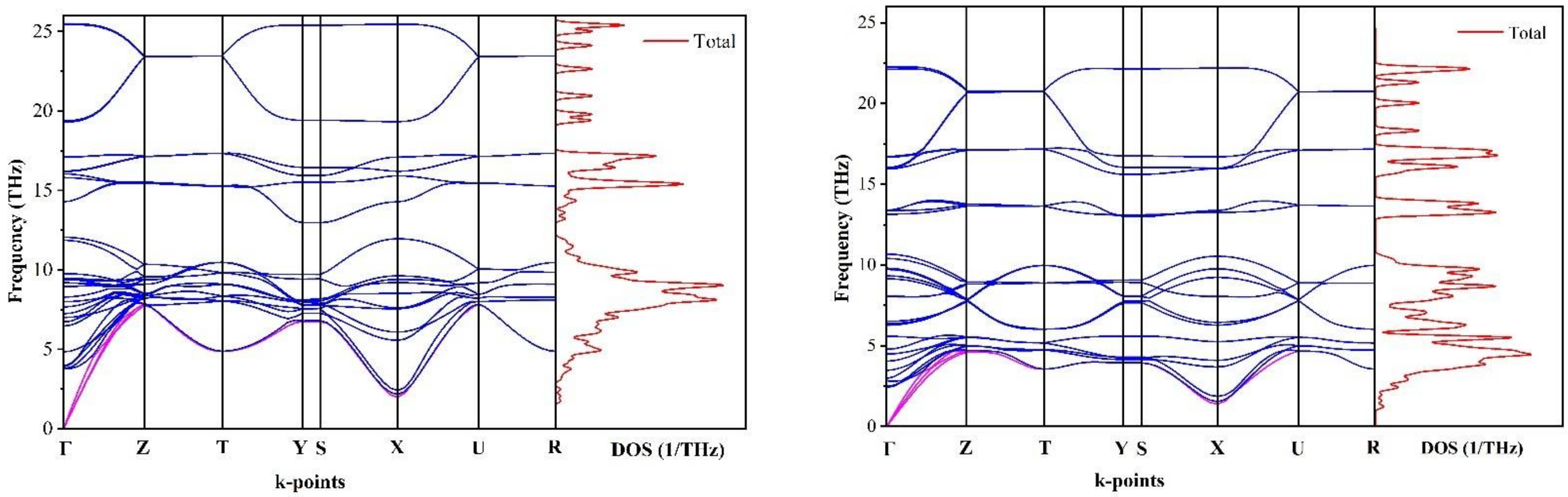

**Fig. 9.** Phonon dispersion spectra (blue and pink line) and total phonon DOS (red line) of (a) VAlB compound and (b) TaAlB compound at zero pressure.

3.8. Thermal properties

3.8.1. Debye temperature

The Debye model is important for understanding the thermal characteristics of a solid that are connected to lattice vibration. It also provides a simple yet efficient way for describing the phonon contributions to the Gibbs energy of crystalline phases. Furthermore, the Debye temperature distinguishes between classical and quantum mechanical phonon activity and aids in the separation of high- and low-temperature behavior zones in solids. The thermal characteristics of crystal lattices, including thermal expansion, thermal conductivity, isothermal compressibility, melting point, heat capacity, and lattice enthalpy, may be well linked by a single parameter: the Debye temperature. At temperatures above $\theta_D$, all modes of vibration have the same energy, $\sim k_B T$. However, higher frequency modes are considered frozen for temperatures below $\theta_D$ [126]. At low temperatures, acoustic vibrations are the dominant vibrational excitations in crystals. At low temperatures, the values of Debye temperature derived using elastic constants and specific heat measurements concur. We computed the Debye temperature of our compounds using the Anderson model, which is an established approach [127]:

$$\theta_D = \frac{h}{k_B}\left[\left(\frac{3n}{4\pi V_0}\right)\right]^{\frac{1}{3}} v_m \tag{45}$$

where $h$ is Planck's constant, $k_B$ is Boltzmann's constant, $n$ is the number of atoms in the unit cell, $V_0$ is the equilibrium volume of the unit cell, and $v_m$ is the average sound velocity in the solid. The equation above shows that $\theta_D$ is proportional to the average sound velocity ($v_m$), which is determined by a crystal's elastic characteristics. The following equations can be used to determine $v_m$ based on the bulk ($B$) and shear ($G$) modulus and the longitudinal ($v_l$) and transverse ($v_t$) sound speeds [128]:

$$v_m = \left[\frac{1}{3}\left(\frac{2}{v_t^3} + \frac{1}{v_l^3}\right)\right]^{-\frac{1}{3}} \tag{46}$$

where,

$$v_t = \sqrt{\frac{G}{\rho}} \tag{47}$$

$$v_l = \sqrt{\frac{3B+4G}{3\rho}} \tag{48}$$

$\rho$ represents the mass density. Table 11 lists the computed Debye temperatures of *M*AlB (*M* = V, Ta, Mo, and Nb) compounds at 0 K. Table 11 shows that VAlB has much higher $v_m$ and $\theta_D$ values than other MAB compounds. This is owing to the decreased mass density of VAlB compound compared to the other compounds. A lower Debye temperature corresponds to poorer phonon thermal conductivity. Hence TaAlB is projected to have lower phonon thermal conductivity, making it a better option for thermal barrier coating than the other MAB compounds under investigation. The Debye temperature also represents the bonding strength in crystals. The Debye temperature of VAlB is high, indicating that the bonding strengths between the atoms are stronger and that the compound is harder. The bonding strength order is as follows: VAlB > MoAlB > NbAlB > TaAlB. Table 11 further indicates that for all four materials, longitudinal sound speeds are larger than transverse sound speeds.

**Table 11:** Calculated crystal density ($\rho$ in g/cm³), longitudinal, transverse and average sound velocities ($v_t$, $v_l$ and $v_m$ in km/s) and Debye temperature ($\theta_D$ in K) of $M$AlB ($M$ = V, Ta, Mo, Nb) compounds.

| **Compound** | $\boldsymbol{\rho}$ | $\boldsymbol{v_l}$ | $\boldsymbol{v_t}$ | $\boldsymbol{v_m}$ | $\boldsymbol{\theta_D}$ | **Ref.** |
|---|---|---|---|---|---|---|
| VAlB | 4.46 | 8.67 | 5.34 | 6.79 | 908 | This work |
| TaAlB | 9.42 | 6.18 | 3.69 | 4.74 | 603 | This work |
| MoAlB | 6.33 | 7.95 | 4.77 | 5.28 | 693 | [34] |
| NbAlB | 5.67 | 7.84 | 4.71 | 5.21 | 664 | [35] |

3.8.2. Melting temperature

The melting temperature ($T_m$) is an important thermophysical characteristic that determines a material's suitability at high temperatures. Crystals with greater melting temperatures have stronger atomic interactions, higher cohesive and bonding energies, and a lower coefficient of thermal expansion. Solids can be utilized continuously at temperatures below $T_m$ without oxidation, chemical change, or severe distortion, which would result in mechanical/elastic failure. The melting temperature of $M$AlB ($M$ = V, Ta, Mo, Nb) compounds has been determined from the elastic constants using the following formula [129]:

$$T_m = 354 + 1.5(2C_{11} + C_{33}) \tag{49}$$

Table 14 shows the predicted $T_m$ values. The $M$AlB ($M$ = V, Ta, Mo, Nb) compounds have melting temperatures of 1780.63 K, 1722.21 K, 2000.70 K, and 1690.50 K, respectively. A crystal's bonding strength is closely related to its lattice (phonon) thermal conductivity, $k_{ph}$, and $T_m$. MoAlB has a high melting temperature because its elastic constants are relatively high when compared to other MAB compounds. Thus, MoAlB will have stronger bonding and cohesive energy than the other three MAB compounds. All of these MAB compounds are ideal for high-temperature applications. TBC materials should have a high melting point, allowing them to tolerate high working temperatures without melting. Stiffer materials have higher melting points. A high melting point is mostly caused by a high heat of fusion, a low entropy of fusion, or a combination of both.

3.8.3. Anisotropies in acoustic velocity

The acoustic velocity of a solid is an important characteristic that is related to its thermal and electrical characteristics. A solid's atoms generate three vibrational modes: two transverse ($\upsilon_{t1}$ and $\upsilon_{t2}$) and one longitudinal ($\upsilon_l$) mode. When sound waves travel through a solid, they stimulate these vibrational modes. The presence of elastic anisotropy in a solid may be identified using sound velocities in different directions. In anisotropic crystals, pure longitudinal and transverse vibration modes can only be detected in particular crystallographic directions, although quasi-transverse or quasi-longitudinal modes can be found in all directions. Because VAlB and TaAlB compounds are an orthorhombic system, the acoustic velocities along the [100], [010], and [001] directions are predicted to be different and may be estimated using the formulae provided below [130]:

$$[100]\upsilon_l = \sqrt{\frac{C_{11}}{\rho}}; \qquad [010]\upsilon_{t1} = \sqrt{\frac{C_{66}}{\rho}}; \qquad [001]\upsilon_{t2} = \sqrt{\frac{C_{55}}{\rho}} \qquad (50)$$

$$[010]\upsilon_l = \sqrt{\frac{C_{22}}{\rho}}; \qquad [100]\upsilon_{t1} = \sqrt{\frac{C_{66}}{\rho}}; \qquad [001]\upsilon_{t2} = \sqrt{\frac{C_{44}}{\rho}} \qquad (51)$$

$$[001]\upsilon_l = \sqrt{\frac{C_{33}}{\rho}}; \qquad [100]\upsilon_{t1} = \sqrt{\frac{C_{55}}{\rho}}; \qquad [010]\upsilon_{t2} = \sqrt{\frac{C_{44}}{\rho}} \qquad (52)$$

The calculated sound velocities along different crystallographic axes of VAlB and TaAlB compounds are presented in Table 12.

**Table 12:** Anisotropic sound velocities ($ms^{-1}$) of VAlB and TaAlB compounds along different crystallographic directions.

| **Propagation direction** | | **VAlB** | **TaAlB** |
|---|---|---|---|
| [100] | $[100]\upsilon_l$ | 8328.45 | 5485.26 |
| | $[010]\upsilon_{t1}$ | 6284.72 | 4344.62 |
| | $[001]\upsilon_{t2}$ | 6207.91 | 4328.83 |
| [010] | $[010]\upsilon_l$ | 7134.83 | 5553.54 |
| | $[100]\upsilon_{t1}$ | 6284.72 | 4344.62 |
| | $[001]\upsilon_{t2}$ | 5704.61 | 4070.50 |
| [001] | $[001]\upsilon_l$ | 8632.63 | 6054.25 |
| | $[100]\upsilon_{t1}$ | 6207.91 | 4328.83 |
| | $[010]\upsilon_{t2}$ | 5704.61 | 4070.50 |

Table 12 demonstrates that the sound velocities of VAlB and TaAlB compounds are not the same in the [100], [010], and [001] directions, implying that the materials will exhibit lattice dynamical anisotropy. Thus, we may predict that VAlB and TaAlB will have direction-dependent thermal and charge transport features. VAlB exhibits much larger sound velocities in all three directions compared to TaAlB.

3.8.4. Lattice thermal conductivity

The lattice (phonon) thermal conductivity ($k_{ph}$) is a crucial parameter in a wide range of vital technologies, including the creation of novel thermoelectric materials, heat sinks, and thermal barrier coating [116,127,131,132]. The lattice thermal conductivity ($k_{ph}$) is a measure of a material's capacity to transmit heat via phonons. It shows how heat is efficiently carried through a material as a result of phonon propagation. High thermal conductivity materials are ideal for effective heat dissipation in applications such as electronic devices, whilst low thermal conductivity materials are beneficial as thermal barriers. In this investigation, we estimated the $k_{ph}$ of VAlB and TaAlB compounds at ambient temperature (300 K) using the following empirical formula published by Slack [133] and the obtained values are disclosed in Table 13.

$$k_{ph} = A \frac{M_{av} \Theta_D^3 \delta}{\gamma^2 n^{2/3} T} \qquad (53)$$

In the equation above, $M_{av}$ is the average atomic mass in kg/mol, $\theta_D$ is the Debye temperature in K, $\delta$ is the cubic root of average atomic volume in meter (m), $n$ is the number of atoms in the conventional unit cell, $T$ is the absolute temperature in K, and $\gamma$ is the acoustic Grüneisen parameter that measures the degree of anharmonicity of the phonons. A material with a low Grüneisen parameter value exhibits low phonon anharmonicity, resulting in high thermal conductivity. The following equation may be used to determine this dimensionless Grüneisen parameter based on the Poisson's ratio [134]:

$$\gamma = \frac{3(1+\sigma)}{2(2-3\sigma)} \qquad (54)$$

The factor A($\gamma$), due to Julian [135], is computed from:

$$A(\gamma) = \frac{5.720\times10^7\times0.849}{2\times(1-0.514/\gamma+0.228/\gamma^2)} \qquad (55)$$

Table 13 shows the calculated lattice thermal conductivity at room temperature (300 K), as well as the Grüneisen value. A high value of the Grüneisen parameter leads to high crystal compressibility and coefficient of thermal expansion. The computed values of the Grüneisen parameter for the MAB phases are low. Among the four VAlB has the lowest phonon anharmonicity, low crystal compressibility and low thermal expansion than other studied compounds. Many of the MAX phase compounds have comparable modest Grüneisen parameters [136,137], indicating moderate lattice anharmonicity. VAlB has a higher Debye temperature ($\theta_D$ = 908 K), which coincides with its high lattice thermal conductivity.

3.8.5. Minimum thermal conductivity and its anisotropy

The behavior of a solid at temperatures above the Debye temperature has become a concern in high temperature applications. At high temperatures, a compound's intrinsic thermal conductivity approaches its lowest value, known as the minimum thermal conductivity ($k_{min}$). This characteristic is crucial because it is unaffected by the presence of impurities or defects in the crystal. Clarke used the quasi-harmonic Debye model to propose the following equation for predicting the $k_{min}$ at high temperatures [138]:

$$k_{min} = k_B \upsilon_m (V_{\text{atomic}})^{-\frac{2}{3}} \qquad (56)$$

In this equation, $k_B$, $\upsilon_m$ and $V_{\text{atomic}}$ represent the Boltzmann constant, average sound velocity, and cell volume per atom, respectively. Materials with higher sound velocity, lowest phonon mean free path, and Debye temperature exhibit higher minimum thermal conductivity. Table 13 shows the estimated isotropic minimum thermal conductivity values for VAlB and TaAlB.

Materials having anisotropic thermal conductivity have a wide range of applications, including heat spreading in electrical and optical device technologies, as well as heat shields, thermoelectric and thermal barrier coatings. Apart from layered composites, elastically anisotropic materials provide the best prospects for intrinsic anisotropic thermal conductivity since heat transport is dominated by elastic wave propagation. There are three mechanisms of heat transmission through solids: thermal vibrations of atoms, movement of free electrons in metals and radiation. To discuss the anisotropy of thermal conductivity, the lowest thermal conductivities of VAlB and TaAlB compounds along distinct crystal orientations are examined using the Cahill's model [139]:

$$k_{min} = \frac{k_B}{2.48} n^{2/3} (\upsilon_l + \upsilon_{t1} + \upsilon_{t2}) \tag{57}$$

and $n = N/V$, where $k_B$ is the Boltzmann constant, $n$ is the number of atoms per unit volume, and $N$ is the total number of atoms in the cell with volume $V$.

**Table 13:** Grüneisen parameter $(\gamma)$, The number of atoms per mole of the compound ($n$ in m$^{-3}$), lattice thermal conductivity ($k_{ph}$ in W/m. K) at 300 K, and minimum thermal conductivity ($k_{min}$ in W/m. K) of *M*AlB (*M* = V, Ta, Mo, Nb) compounds.

| **Compound** | $\gamma$ | ***n*** **($10^{28}$)** | $k_{ph}$ | **[100]**$k_{min}$ | **[010]**$k_{min}$ | **[001]**$k_{min}$ | $k_{min}$ | **Ref.** |
|---|---|---|---|---|---|---|---|---|
| VAlB | 1.23 | 9.06 | 68.84 | 2.34 | 2.14 | 2.31 | 1.89 | This work |
| TaAlB | 1.36 | 7.78 | 42.13 | 1.44 | 1.42 | 1.47 | 1.19 | This work |
| MoAlB | 1.36 | - | - | - | - | - | - | [34] |
| NbAlB | 1.36 | - | - | - | - | - | - | [35] |

Table 13 shows the estimated $k_{min}$ values of VAlB and TaAlB for the crystallographic orientations [100], [010], and [001]. From Table 13 it is seen that the Clarke model predicts a lower minimum thermal conductivity than the Cahill model. It is also observed that the $k_{min}$ of VAlB is higher than that for TaAlB. It can also be concluded from the above table that for VAlB, the $k_{min}$ along [100] direction is slightly higher than that along [010] and [001] direction and for TaAlB compound, $k_{min}$ along [001] direction is slightly higher than that along [100] and [010] direction. This implies about the existence of small anisotropy in thermal transport for both compounds.

3.8.6. Thermal expansion coefficient, heat capacity and wavelength of the dominant phonon mode

Thermal expansion behavior is an inherent thermal feature caused by anharmonic lattice vibrations. The thermal expansion coefficient affects several properties of a substance, including thermal conductivity, specific heat, and entropy. Materials with very low thermal expansion are not only useful in practice, but also of basic scientific importance. To compute a material's thermal expansion coefficient ($\alpha$) from its shear modulus, we use the following equation [100]:

$$\alpha = \frac{1.6 \times 10^{-3}}{G} \qquad (58)$$

where, $G$ is the isothermal share modulus in GPa. The thermal expansion coefficient of a material is inversely linked to its melting temperature ($\alpha \approx 0.02/T_m$) [138,140]. Crystals' thermal expansion coefficients vary with temperature. Table 14 shows the computed values of VAlB and TaAlB compounds at 300 K. Both VAlB and TaAlB structures have exceptionally low thermal expansion coefficients. Low thermal expansion materials are widely used in high anti-thermal shock applications (for example, cookware for use in an oven or freezer), electronic devices, heat-engine components, spintronic devices, and so on.

The dominant phonon wavelength ($\lambda_{dom}$) is where the phonon distribution reaches its maximum. The $\lambda_{dom}$ affects thermal and electrical transmission in materials. Long-wavelength acoustic phonons have a larger role in heat transfer as temperatures fall. The wavelength of the dominating phonon for materials at different temperatures may be determined using the following expression [138,140]:

$$\lambda_{dom} = \frac{12.566\, v_m}{T} \times 10^{-12} \qquad (59)$$

where, $v_m$ is the average sound velocity in $ms^{-1}$, and $T$ is the temperature in K. Materials having greater average sound velocity, higher shear modulus, and lower density have longer dominant phonon wavelength. Table 14 summarizes the expected values for $\lambda_{dom}$ at room temperature.

The heat capacity ($C_P$) is a key thermodynamic quantity in the material design process. Materials with larger heat capacity have higher thermal conductivity ($k$) and lower thermal diffusivity (D), as $k = \rho C_P D$. The volumetric heat capacity of a material is defined as the change in thermal energy per unit volume for each Kelvin of temperature change. The volumetric heat capacity of a substance may be determined using the following empirical formula [100,140]:

$$\rho C_P = \frac{3k_B}{\Omega} \qquad (60)$$

where, $N = 1/\Omega$ is the number of atoms per unit volume. The heat capacity per unit volume of VAlB and TaAlB is given in Table 14.

**Table 14:** Calculated melting temperature ($T_m$ in K), thermal expansion coefficient ($\alpha$ in $K^{-1}$), heat capacity per unit volume ($\rho C_\rho$ in $JK^{-1}m^{-3}$), and wavelength of the dominant phonon mode ($\lambda_{dom}$ in m) at 300 K of *M*AlB (*M* = V, Ta, Mo, Nb) compounds.

| Compound | $T_m$ | $\alpha$ ($10^{-5}$) | $\rho C_\rho$ ($10^{6}$) | $\lambda_{dom}$ ($10^{-12}$) | Ref. |
|---|---|---|---|---|---|
| VAlB | 1780.63 | 1.25 | 3.75 | 284.41 | This work |
| TaAlB | 1722.21 | 1.24 | 3.22 | 198.54 | This work |
| MoAlB | 2000.70 | - | - | - | [34] |
| NbAlB | 1690.50 | - | - | - | [35] |

3.9. Optical properties

The optical study of solids has sparked widespread interest in modern science and technology. Optical materials are widely used in display devices, lasers, photodetectors, solar cells, sensors, and reconfigurable photonics [141]. The interaction of a substance with incoming electromagnetic radiation is determined by its optical characteristics. The major focus of optoelectronic devices is on their reaction to visible light. Such investigations forecast which materials will be suitable for use in various optoelectronic devices. The energy-dependent optical properties often examined include dielectric function $\varepsilon(\omega)$, refractive index $n(\omega)$, optical conductivity $\sigma(\omega)$, reflectivity $R(\omega)$, absorption coefficient $\alpha(\omega)$, and energy loss function $L(\omega)$ (where $\omega = 2\pi f$ is the angular frequency of the electromagnetic wave). The response of these parameters to incident photon energy/frequency has been estimated and investigated in this section. The complex dielectric function is written as follows:

$$\varepsilon(\omega) = \epsilon_1(\omega) + i\epsilon_2(\omega) \quad (61)$$

Kramers-Kronig relationships connect the two parts. All additional optical constants of importance are given by the following relationships [142]:

$$n(\omega) = \sqrt{\frac{|\varepsilon(\omega)| + \epsilon_1(\omega)}{2}} \quad (62)$$

$$k(\omega) = \sqrt{\frac{|\varepsilon(\omega)| - \epsilon_1(\omega)}{2}} \quad (63)$$

$$R(\omega) = \frac{(n-1)^2 + k^2}{(n+1)^2 + k^2} \quad (64)$$

$$\alpha(\omega) = \frac{2k\omega}{c} \quad (65)$$

$$L(\omega) = Im\left(\frac{-1}{\varepsilon(\omega)}\right) = \frac{\epsilon_2(\omega)}{\epsilon_1^2(\omega) + \epsilon_2^2(\omega)} \quad (66)$$

$$\sigma(\omega) = \sigma_1(\omega) + i\sigma_2(\omega) \quad (67)$$

In this part, we have only shown the results for VAlB and TaAlB compounds when incident electric fields are in the [100], [010], and [001] directions. Comprehensive theoretical study of optical parameters of MoAlB and NbAlB exist in the literature [35,143]. Therefore, comparison of the optical properties of $M$AlB ($M$ = V, Ta, Mo, Nb) compounds can be made. The optical parameters of VAlB and TaAlB compounds are depicted in Figs. 10 and 11 for incident energy up to 30 eV and the electric field polarizations along the three principle crystallographic directions. The real and imaginary parts of the dielectric function thoroughly illuminate a variety of optical features of materials, including excitons, free carrier absorption, superconducting gaps, plasmon resonances, and intra- and inter-band electronic transitions [144–146]. For metallic materials, the Drude damping adjustment is necessary [131,147]. For all calculations, we have employed 0.5 eV Gaussian smearing. For our computation, we employ an empirical Drude term with plasma frequency of 8 eV and damping of 0.05 eV.

The absorption coefficient $\alpha(\omega)$ provides data about optimum solar energy conversion efficiency and it indicates how far light of a specific energy (wavelength) can penetrate into the material before being absorbed. The absorption spectra of VAlB and TaAlB are shown in Figs. 10(a) and 11(a), respectively, reveal the metallic nature since the spectra starts from 0 eV. In [100] direction, the absorption coefficient is quite high in the UV region and peaks at 8.81 eV and 8.55 eV for VAlB and TaAlB, respectively. Again, in [010] direction, the absorption coefficient peaks at 7.47 eV and 9.69 eV for VAlB and TaAlB, respectively. Furthermore, along [001] direction the absorption coefficient peaks at 12.5 eV and 11.4 eV for VAlB and TaAlB, respectively. We know the range of quantum energies of ultraviolet radiation is 3.1 - 124 eV. So, both the compounds are promising for absorbing materials in the UV region. The absorption coefficient drops down quickly at the plasma edge, at around 18 eV for both the compounds. The absorption coefficient exhibits a moderate level of optical anisotropy.

Optical conductivity $\sigma(\omega)$ provides an explanation for the conduction of free charge carriers over a certain range of photon energy. As can be seen from the band structure, the materials have no band gap, hence photoconductivity begins at zero photon energy, as Figs. 10(b) and 11(b) illustrates. The optical conductivity peaks exhibit direction dependence of the electric field polarization. The photoconductivity and hence electrical conductivity of the materials increase as a result of absorbing photons.

The dielectric function $\varepsilon(\omega)$ shows how a material interacts with the incident electromagnetic waves. The dielectric function is closely related to the electronic band structure. The real $\epsilon_1(\omega)$ and imaginary part $\epsilon_2(\omega)$ of the dielectric function for VAlB and TaAlB along [100], [010] and [001] directions are shown in Figs. 10(c) and 11(c). The dielectric function's imaginary part is connected to the dissipation of electromagnetic wave energy inside the medium, whereas the real part is related to electrical polarization and anomalous dispersion. The compounds have metallic properties within the energy ranges where, $\epsilon_1(\omega) < 0$. The findings indicate that MoAlB exhibits Drude-like behavior and is a suitable dielectric material [143]. In the energy ranges, MoAlB and NbAlB's dielectric functions also show metallic features [35,143]. Optical anisotropy is seen in all MAB compounds.

In the [100] direction $\epsilon_1(\omega)$ and $\epsilon_2(\omega)$ approaches zero at 22.62 eV and 23.42 eV for VAlB and TaAlB, respectively. For the real part $\epsilon_1(\omega)$ of dielectric function, the peaks are around 2.67 eV and 2.42 eV for VAlB and TaAlB, respectively in the same direction. However, in the [010] the two part of the dielectric function approaches zero at 22.24 eV and 24.21 eV for VAlB and TaAlB, respectively. In this direction the peaks for $\epsilon_1(\omega)$ are around 2.58 eV and 2.17 eV for VAlB and TaAlB, respectively. Moreover, in the [001] direction the real and imaginary part of the dielectric function approaches zero at 21.62 eV and 23.95 eV for VAlB and TaAlB, respectively and the peaks for $\epsilon_1(\omega)$ around 2.47 eV and 2.17 eV for VAlB and TaAlB, respectively. In the high energy zone, the magnitudes of both $\epsilon_1(\omega)$ and $\epsilon_2(\omega)$ decline monotonically. For both compounds, the imaginary portion approaches zero at about 23 eV. The peaks in the $\epsilon_2(\omega)$ originate from the combined impacts of matrix elements of photon driven electronic transitions between electronic states and their corresponding energy density of states. Regarding electric field polarization, the dielectric function exhibits moderate optical anisotropy.

The loss function $L(\omega)$, for VAlB and TaAlB phases under study is displayed in Figs. 10(d) and 11(d) respectively. It describes the energy loss of a fast electron traversing in the material. In the energy-loss spectrum, we observed that the highest peaks of VAlB compound are at 22.32 eV, 23.12 eV and 22.35 eV along [100], [010] and [001] polarization directions, respectively. For TaAlB the highest peaks are located at 23.11 eV, 23.83 eV and 23.32 eV along [100], [010] and [001] directions, respectively. So, both compounds are optically anisotropic. The plasma resonance is linked to the peaks in energy loss function spectra, and the related frequency is known as the plasma frequency ($\omega_p$) [132]. The effective mass and concentration of the electrons determine the frequency of collective oscillations that the charge carriers experience at this specific energy. The high energy region's energy loss peak may be seen at $\epsilon_2 < 1$ and $\epsilon_1 = 0$ [51]. The trailing edges in the absorption and reflection spectra are matched by the sharp energy loss peaks [148,149]. The plasma oscillations due to collective motions of the charge carriers are induced at these particular energies. When the incident photon frequency is higher than plasma frequency ($\omega_p$), the material becomes transparent to incident light.

Relative to the energy of the wave striking a surface, the reflectivity $R(\omega)$ is the ratio of the wave's energy reflected from the surface. Figs. 10(e) and 11(e), respectively, display the reflectivity spectra of the compounds VAlB and TaAlB as a function of photon energy. STARTFor VAlB, it is observed from Fig. 10(e) that the reflectivity, starts with a value 98%, decreases sharply at 1.88 eV, 2.63 eV and 1.96 eV along the polarization directions [100], [010] and [001] respectively. Again, start increasing and reach maximum value at 9.56 eV, 10.41 eV and 16.42 eV for directions [100], [010] and [001] respectively. For TaAlB, we observed from Fig. 11(e) that the reflectivity, starts with a value of 99% decreases sharply up to 2.05 eV, 1.65 eV and 2.14 eV along [100], [010] and [001] directions respectively. Then starts increasing and reach maximum value at 17.21 eV, 10.3 eV and 16.42 eV along directions [100], [010] and [001], respectively. The infrared spectrum has extremely high reflectance. The visible to ultraviolet light (2 eV – 20 eV) portion of the reflectivity spectrum is high and non-selective. This is a desirable optical property of TaAlB and VAlB, and the MAB phases have a great deal of promise for use as a very effective solar radiation reflector. Like the absorption coefficient, the reflectivity of VAlB and TaAlB compounds falls dramatically at energies over 20 eV. The anisotropy in reflectivity is low.

Figs. 10(f) and 11(f), respectively, show the refractive index, $n$ (real part), and extinction coefficients, $k$ (imaginary part), of VAlB and TaAlB compounds in the [100], [010], and [001] directions. The imaginary portion of the refractive index, $k$, indicates the attenuation of the electromagnetic wave as it passes through a medium, whereas the real part depicts the phase velocity of the electromagnetic wave at different energies. The near ultraviolet (UV) region is the peak for the extinction coefficient, $k$, for VAlB and TaAlB, respectively, at 3.89 eV and 3.69 eV in all three polarization directions. $n$ (real part) has high values in the infra-red and visible regions. Solids with a high refractive index can be used as wave guides. Refractive index optical anisotropy is rather small.

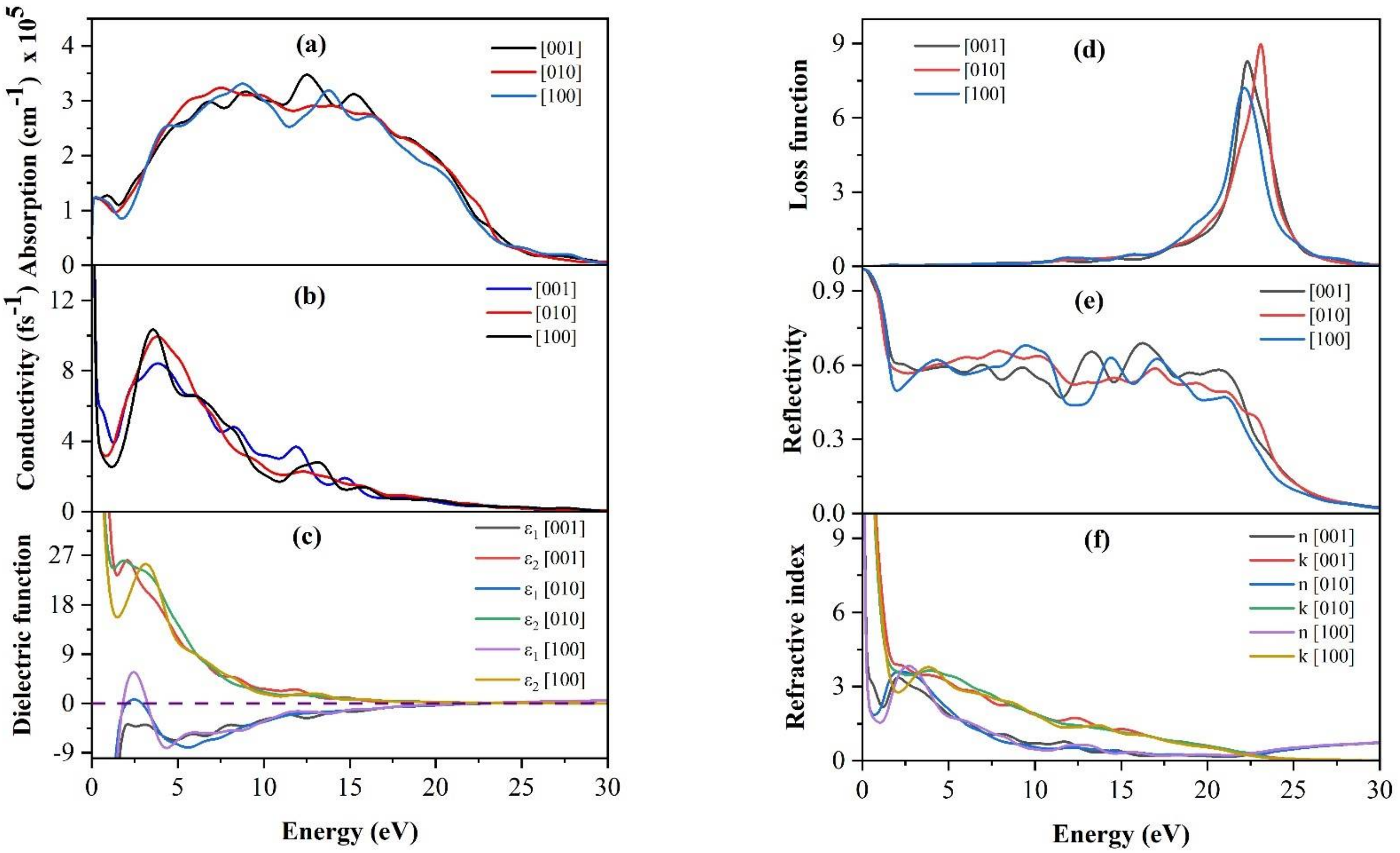

**Fig. 10.** The frequency-dependent (a) absorption coefficient (b) optical conductivity (c) dielectric function (d) loss function (e) reflectivity, and (f) refractive index of VAlB with electric field polarization vectors along [100], [010] and [001] directions.

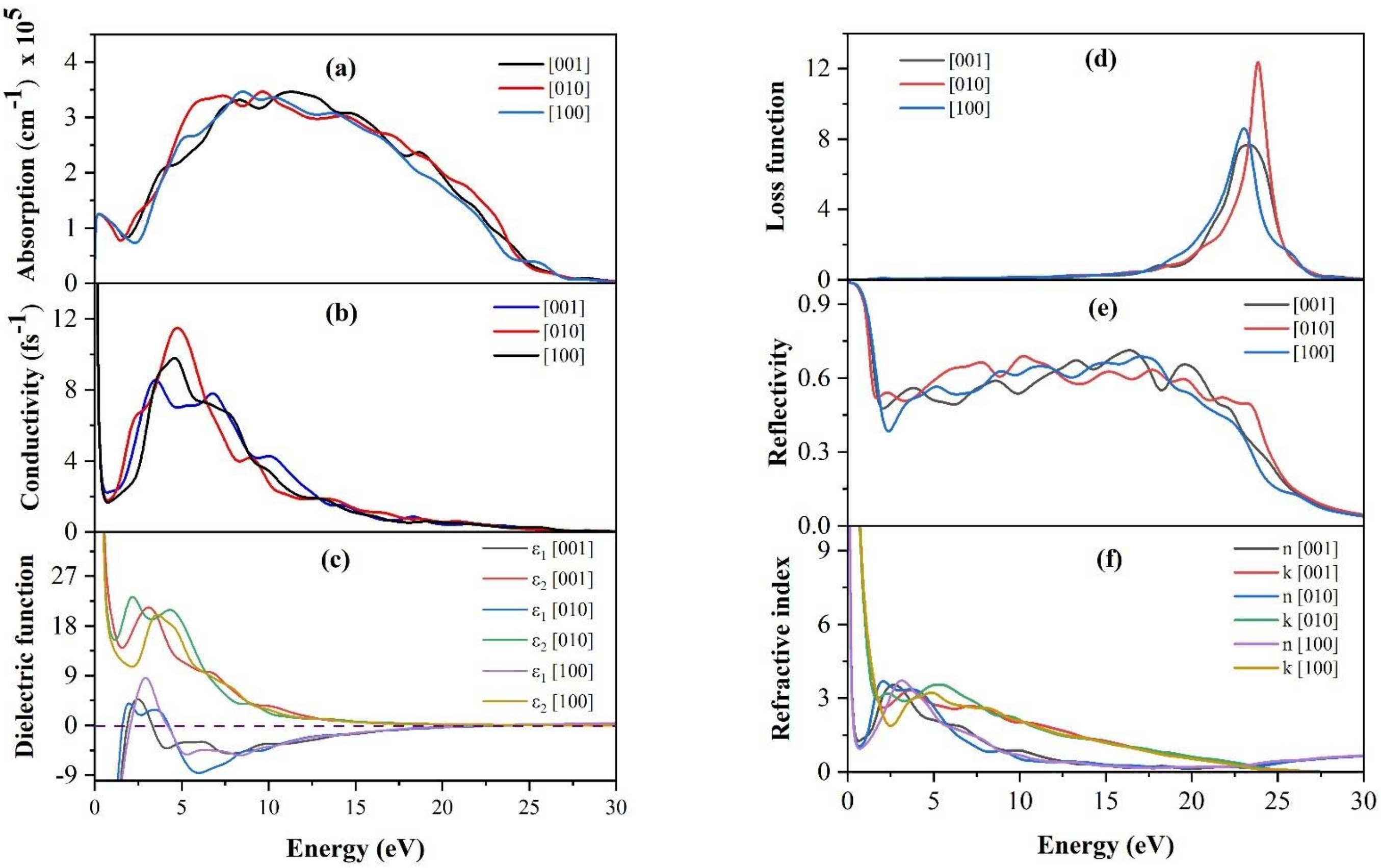

**Fig. 11.** The frequency-dependent (a) absorption coefficient (b) optical conductivity (c) dielectric function (d) loss function (e) reflectivity and (f) refractive index of TaAlB with electric field polarization vectors along [100], [010] and [001] directions.

## 4. Conclusions

In this work, we explored two new members of the technologically prominent MAB phases VAlB and TaAlB, utilizing DFT-based calculations and an extensive comparison to previously investigated MAB compounds MoAlB and NbAlB [34,35] has been made.

The estimated lattice parameters of VAlB and TaAlB accord well with the experimental results. These materials have been shown to be chemically, mechanically, and dynamically stable. Among these MAB compounds, NbAlB exhibits the highest overall elastic anisotropy. The bulk modulus to shear modulus ratio, Poison's ratio, and Cauchy pressure analysis all indicate that VAlB and TaAlB, like MoAlB and NbAlB compounds, should be brittle. All of these MAB phase compounds are relatively hard and fairly machinable, although VAlB is the hardest and TaAlB is the most machinable, making them desirable compounds for engineering purposes. NbAlB has the highest fracture toughness of any of the compounds studied in this research. The VAlB and TaAlB compounds have low compression anisotropy but high shear anisotropy. The charge density mapping, MPA, and HPA findings show that metallic populations for *M*AlB (*M* = V, Ta, Mo, Nb) compounds are low, indicating the presence of weak metallic bonds in these compounds. Ionic and covalent bonds dominate in these compounds. The band structure of *M*AlB (*M* = V, Ta, Mo, Nb) compounds demonstrates their metallic character, since there is a considerable overlap of conduction and valence bands with variable degrees of dispersion at the Fermi level. There is strong hybridization between the V 3*d*, Al 3*p*, and B 2*p* electronic states of the VAlB and the 5*d*, Al 3*p*, and B 2*p* electronic states of the TaAlB at the Fermi energy. The TDOS of the MAB compound VAlB is much greater, indicating that its electrical conductivity is much higher than that of other MAB compounds. The calculated Coulomb pseudopotential of TaAlB is less than that of VAlB, implying that electronic correlations are larger in VAlB. The resulting Fermi surface topology is complex, containing both electron and hole-like sheets. The predicted elastic moduli, Debye temperature, minimum thermal conductivity, and melting temperature indicate that all the compounds have the potential to be employed as thermal barrier coating (TBC) materials. The relatively high Debye temperature of VAlB suggests a strong thermal conductivity than the other investigated MAB compounds. The acoustic velocities in the MAB phase compounds are comparable. MoAlB has a substantially higher melting temperature than other MAB compounds, indicating a wide variety of potential high temperature applications. The optical characteristics of VAlB and TaAlB are studied in depth and compared to prior studies. All of the optical parameters exhibit metallic properties. Each of these substances exhibits optical anisotropy. These materials have been discovered to be excellent UV radiation absorbers. The reflectance spectrum is nonselective throughout a broad energy range. These MAB compounds have strong reflective properties, making them ideal for decreasing solar heating. The high value of the real component of the refractive index in the infrared and visible range indicates that these MAB compounds can be employed in optoelectronic device applications.

To summarize, this study theoretically investigates a wide range of physical properties of *M*AlB (*M* = V, Ta, Mo, Nb) compounds. All the compounds have various appealing mechanical, thermal, and optoelectronic properties that make them ideal for engineering and optical device applications. We hope that the findings presented here will encourage researchers to examine these ternary borides in further depth in the near future, both theoretically and empirically. Moreover, all of the examined physical properties of VAlB and TaAlB are unique and can be utilized as references for future study.

**Acknowledgments**

S.H.N. acknowledges the research grant (1151/5/52/RU/Science-07/19-20) from the Faculty of Science, University of Rajshahi, Bangladesh, which partly supported this work. M.E.H. gratefully acknowledges the National Science and Technology (NST) Fellowship, awarded by the Ministry of Science and Technology, Bangladesh, for supporting his M.Sc. research. This work is dedicated to the cherished memory of the martyrs of the July-August 2024 revolution in Bangladesh, whose sacrifices will forever inspire us.

**Data Availability**

The data sets generated and/or analyzed in this study are available from the corresponding author upon reasonable request.

**Declaration of interest**

The authors declare that they have no known competing financial interests or personal relationships that could have appeared to influence the work reported in this paper.

## Author Contributions

**Jahid Hassan**: Methodology, Software, Writing- Original draft. **M. A. Masum**: Methodology, Software. **Ruman Ali**: Methodology, Software. **Md. Enamul Haque**: Methodology, Software. **S. H. Naqib**: Conceptualization, Supervision, Formal Analysis, Writing- Reviewing and Editing.